\documentclass[final]{ustcstep}

\usepackage[dvips]{graphicx}
\usepackage[square]{natbib}
\usepackage{stfloats}
\usepackage{amssymb}

\def\gsim{\;\lower4pt\hbox{${\buildrel\displaystyle >\over\sim}$}\;}
\def\lsim{\;\lower4pt\hbox{${\buildrel\displaystyle <\over\sim}$}\;}
\def\grls{\;\lower4pt\hbox{${\buildrel\displaystyle >\over <}$}\;}

\newcommand\addr[2]{{\footnotesize \it $^{#1}$#2}\\}

\begin{document}

\title{Quantitative Analysis of CME Deflections in the Corona}

\author{Bin Gui$^{1, 2}$, Chenglong Shen$^1$, Yuming Wang$^{1, *}$, Pinzhong Ye$^{1, 2}$, Jiajia Liu$^1$,\\Shui Wang$^1$, and Xuepu Zhao$^3$ \\[1pt]
\addr{}{$^1$ CAS Key Laboratory of Geospace Environment, Department
of Geophysics $\&$ Planetary Sciences, University of } \addr{}{\,
\,Science $\&$ Technology of China, Hefei, Anhui 230026, China}
\addr{}{$^2$ State Key Laboratory of Space Weather, Center for Space
Science and Applied Research, Chinese Academy of }\addr{}{\,
\,Sciences, Beijing, China} \addr{}{$^3$ W.W. Hansen Experimental
Physics Laboratory, Stanford University, Stanford, CA 94305,
 USA}
\addr{}{$^*$ Author for correspondence, Email: ymwang@ustc.edu.cn}}

\maketitle
\tableofcontents

\begin{abstract}
In this paper, ten CME events viewed by the STEREO twin spacecraft are analyzed to study the deflections of CMEs during their propagation in the corona. Based on the three-dimensional information of the CMEs derived by the graduated cylindrical shell (GCS) model \citep{Thernisien_etal_2006},  it is found that the propagation directions of eight CMEs had changed. By applying the theoretical method proposed by \citet{Shen_etal_2011}  to all the CMEs, we found that the deflections are consistent, in strength and direction, with the gradient of the magnetic energy density. There is a positive correlation between the deflection rate and the strength of the magnetic energy density gradient and a weak anti-correlation between the deflection rate and the CME speed. Our results suggest that the deflections of CMEs are mainly controlled by the background magnetic field and can be quantitatively described by the magnetic energy density gradient (MEDG) model.
\end{abstract}

%\begin{article}

\section{Introduction}\label{sec:01}

Corona mass ejections (CMEs) are large scale eruptions from the solar surface and act as one of the primary drivers of space weather phenomena, such as geomagnetic storms, solar energetic particle events, etc. The deflections of CMEs, which were first reported by
\citet{MacQueen_etal_1986} in the Skylab epoch (1973-1974),  are one of the factors influencing the geoeffectiveness of CMEs. A statistical study about CME deflections was made by
\citet{Cremades_Bothmer_2004}. They identified the source regions of 124 structured CME events observed by the Large Angle and Spectrometric Coronagraph (LASCO) on board the Solar and Heliospheric Observatory (SOHO) during 1996-2002 and compared the source regions of CMEs with their central position angles (CPAs).
\citet{Cremades_Bothmer_2004} found that there was a systematic deflection by 20 degrees to lower latitudes only in activity-minimum years (1996-1998) and no systematic trend nor deflection during the years 1999-2002. The result was further confirmed by the recent work of \citet{Wang_etal_2011}, in which all the LASCO CMEs during
1997-1998 were examined. \citet{Cremades_etal_2006} found a good correspondence between the deflection of CMEs and the total area of coronal holes (CHs). They suggested that the neighboring CHs affect the outward evolution of CMEs near the Sun and cause such deflections. \citet{Shen_etal_2011} analyzed the deflection of the 8 October 2007 CME in the meridian plane in much more details. They showed strong evidence that the trajectory of the CME was influenced by the background magnetic field, and the CME tends to deflect to the region with lower magnetic energy density.

Note that all the CME deflections studied above are in the latitudinal direction. The CME deflections on a spherical surface, i.e., in both latitudinal and longitudinal directions, still remain unclear due to the presence of projection effect. Even so, the CME deflection in longitude was suggested by some researchers. For example, the longitudinal deflections of CMEs as they propagate in the interplanetary space were first proposed by
\citet{Wang_etal_2004b,Wang_etal_2006a}. Such deflections can explain the asymmetrical east-west distribution of the source locations of the geoeffective halo CMEs \citep{Wang_etal_2002a}. \citet{Gopalswamy_etal_2004,Gopalswamy_etal_2005,Gopalswamy_etal_2009} also suggested that CMEs could be deflected away from the Sun-Earth line by the associated coronal holes. They use such deflections to explain the existence of the `driverless' shocks, which were observed near the Earth but without their drivers, the interplanetary coronal mass ejections (ICME).

Since the successful launch of the Solar TErrestrial RElations Observatory (STEREO) mission \citep{Kaiser_etal_2008}, the three-dimensional (3-D) information of CMEs is more or less revealed in observations with the aid of various reconstruction models \citep[e.g.,][]{Thernisien_etal_2006,Thernisien_etal_2009,Lugaz_etal_2010,Liu_etal_2010a,Liu_etal_2010b}. STEREO consists of two identical satellites. It has provided for the first time the observations of the Sun from dual vantage points. Based on the STEREO observations, some CME events with an obvious deflection in the latitudinal direction have been reported  \citep[e.g.,][]{Kilpua_etal_2009,Shen_etal_2011}, and the possible deflections of CMEs in the ecliptic plane have also been discussed \citep[e.g.,][]{Liu_etal_2010b,Lugaz_etal_2010,Poomvises_etal_2010}.

In this paper, we have comprehensively studied the CME deflections in the corona in both latitudinal and longitudinal directions for ten CME events viewed by the STEREO twin spacecraft. The data and the method we used will be introduced in the next section. In Section \ref{sec:03}, four cases will be selected to show different types of deflections, in which their 3-D trajectories and a comparison between the deflections and the magnetic energy density distributions will be presented. In Section \ref{sec:04}, statistical studies on the deflection and its correlation with the magnetic energy density will be presented. Finally, we will give the conclusions and make some discussion in Section \ref{sec:05}.
\section{Data and Method}\label{sec:02}
\subsection{Three-dimensional information of CMEs}

The observations from the COR1 and the COR2 instruments of the Sun Earth Connection Coronal and Heliospheric Investigation (SECCHI) \citep{Howard_etal_2008} suite on board the STEREO A and B spacecraft are used to learn about the evolutions of CMEs in the corona. The COR1 instruments observe the corona from 1.4-4.0 $R_s$ and the COR2
instruments observe the corona from 2.5-15.0 $R_s$. In this paper, these observations were used to obtain the 3-D information of CME during its propagation in the corona. The observations from the SECCHI/EUVI and the Michelson Doppler Imager (MDI) on board SOHO \citep{Brueckner_etal_1995} are used to identify the CMEs' source regions on the solar surface.

To obtain the 3-D geometry and therefore the trajectory of a CME, the graduated cylindrical shell (GCS) model developed by \citet{Thernisien_etal_2006,Thernisien_etal_2009} was applied to both the projected two-dimensional images from the STEREO-A (STA) and the STEREO-B (STB) spacecraft. In that model, CMEs are assumed to have a flux rope-like structure. The GCS model has nine free parameters (refer to Table 1 of \citet{Thernisien_etal_2006}). Six of them determine the CME's shape projected on the plane of the sky. These parameters, referred to as geometric parameters, are the longitude  `$\phi$' and latitude
`$\theta$', height `$h$' (the height of the legs, or `$h_f$', the
height of the leading edge), aspect ratio `$\kappa$', tilt angle
`$\gamma$' with respect to the equator, and half angular width
`$\alpha$' between the flux rope legs. The other three parameters, specifying the electron density distribution at the shell, are the electron density factor `$N_{\varepsilon}$', Gaussian width `$\sigma_{trailing}$' of the density profile in the interior of the
GCS and Gaussian width `$\sigma_{leading}$' of the density profile
at the exterior of the GCS.

A set of reasonable initial values of the parameters is helpful to get the best fitting of the CME images. Observations of the CME source region on the solar surface were used to constrain the initial values of longitude $\phi$, latitude $\theta$ and tilt angle
$\gamma$ if any. The tilt angle can be estimated according to the CME-associated filament (or the polarity inversion line, PIL, if no filament observed) because it is believed that a CME is a flux rope surrounding its associated filament and standing above the PIL. The rest of the parameters are set by comparing the GCS flux rope to the CME shape observed simultaneously by both STA and STB. In practice, we find that the tilt angle and half angle are insensitive to the fitting results. Therefore we fix them to a certain reasonable value for the whole CME evolution process by trial and error. It should be noted that fixing the tilt angle indicates a CME without rotation, which may not be true for many CME events \citep{Lynch_etal_2009,Mostl_etal_2008,Shiota_etal_2010,Torok_etal_2003,Wang_etal_2006c,Yurchyshyn_etal_2007}. However, we find that the change of the tilt angle will not significantly affect the derived directions of CMEs as long as the GCS flux rope matches the observed CME shape in both the STA and STB images (see the discussion in Section \ref{sec:05}). Thus, in this study a fixed tilt angle is acceptable. Besides, not all the CMEs in our sample have available observations of their source regions. For such events, we just compare the GCS flux rope with the observed CME shape to estimate the parameters.

\subsection{Coronal magnetic field}
It is believed that the magnetic energy is dominant in the corona. Previous studies have suggested that the CME deflection can be qualitatively interpreted as the constraint of the ambient magnetic structure, e.g., coronal holes \citep{Gopalswamy_etal_2004}. Our recent study of the 8 October 2007 CME showed that the behavior of the CME's latitudinal deflection can be quantitatively described by a theoretical method, in which the direction and magnitude of the deflection are well consistent with the gradient (with the conventional minus sign in front) of magnetic energy density, $<-\nabla (\frac{B^2}{2\mu_0})>$, where the angle brackets mean the
average over the region occupied by the CME \citep{Shen_etal_2011}.  In that work, however, the deflection in the latitudinal direction of only one CME was studied. Thus, we will test the method with more CME events to check if it is also applicable to other CMEs and to the deflections in other directions.

\begin{figure*}[htbp]
  \centering
  \includegraphics[width=0.95\hsize]{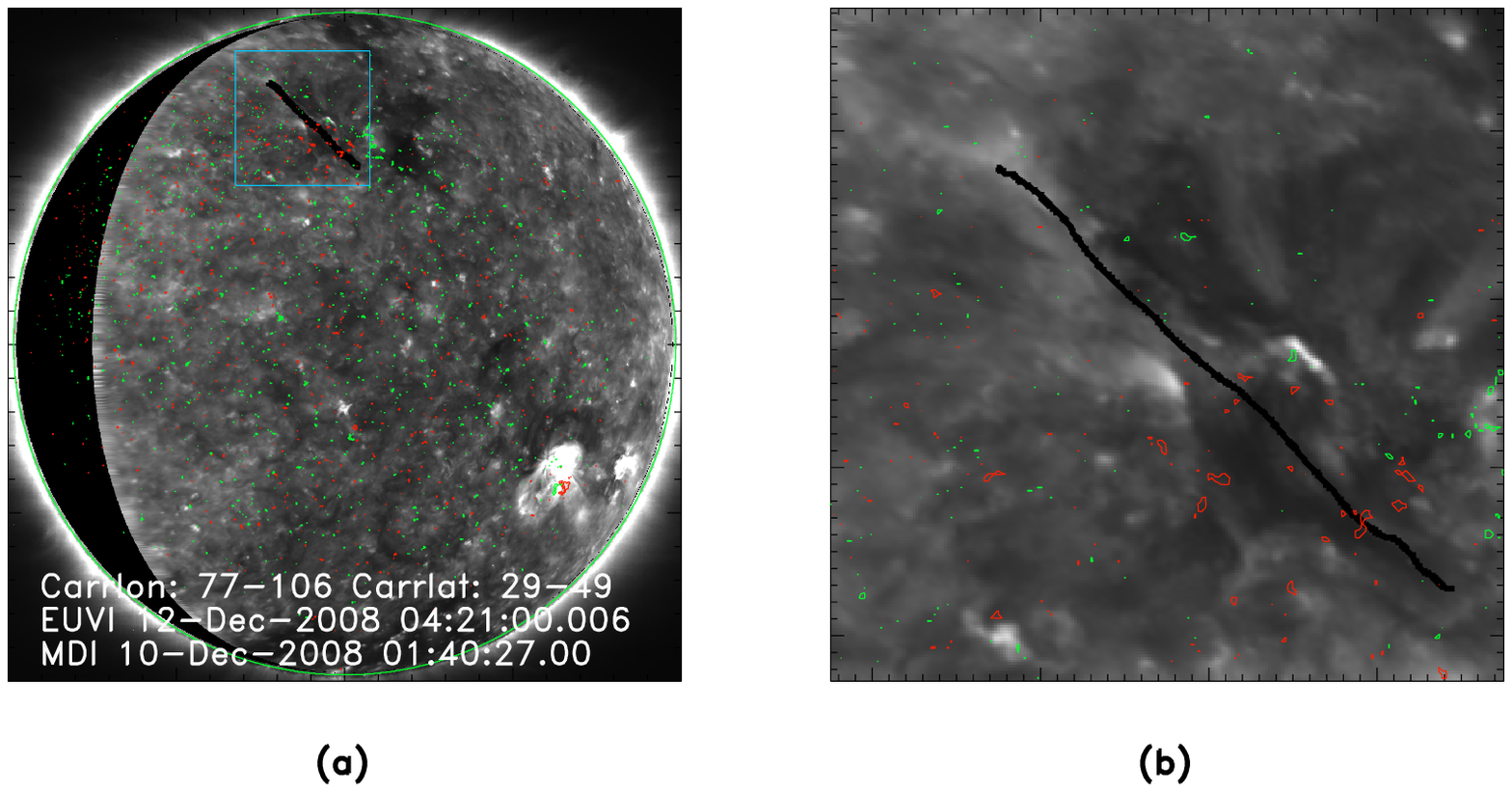}
  \caption{The source region of the 12 December 2008 CME observed by the STEREO/EUVI and the SOHO/MDI. Panel (a) shows the STEREO/EUVI 171 \AA\ image superposed by the contours of the SOHO/MDI magnetogram taken two days before the event when the source region (marked by the square box) was visible to the SOHO. The STEREO/EUVI image was rotated to match the angle of view of the SOHO/MDI. Panel (b) shows the zoomed-in image of the source region.}\label{fg_01}
%\end{figure*}

%\begin{figure*}[t]
%  \centering
  \includegraphics[width=0.95\hsize]{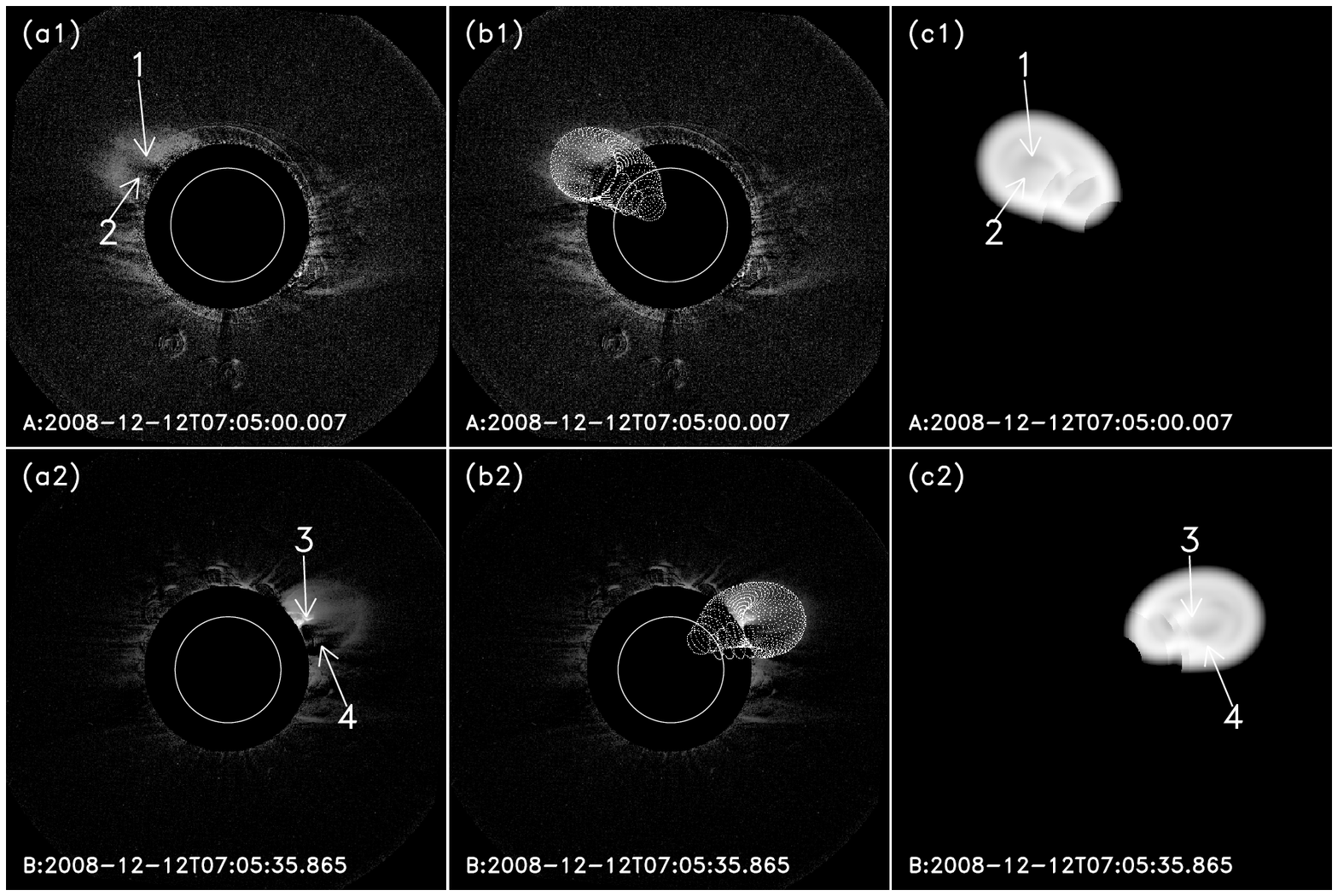}
  \caption{The fitting example of CME-1. From (a) to (c): The original CME images, the modeled wireframe images which overlay on the CME images, and the relative brightness derived from the GCS model. The top and bottom panels present the results based on the STA and STB data, respectively. The arrows with numbers indicate the common points of the CME between the original images and the derived brightness images.}\label{fg_02}
\end{figure*}

In the method, the 3-D magnetic field of the corona is key information, and is extrapolated from the SOHO/MDI photospheric magnetic synoptic charts by the current-sheet source-surface (CSSS) model developed by \citet{Zhao_Hoeksema_1995}. The CSSS model is a development of the potential-field source-surface (PFSS) model, and was used in our previous work \citep{Shen_etal_2007,Shen_etal_2011,Wang_Zhang_2007}. The magnetic synoptic chart is created from the MDI daily magnetograms over a quasi-27-day solar rotation. It cannot reflect the state of the photosphere right at the time of a CME taking place. However, what we are interested in is the large scale coronal magnetic field, which probably changes little during a solar rotation \citep{Ness_Wilcox_1964}. Under this consideration, the synoptic chart may be treated as a good approximation to the real photospheric magnetogram over the full solar surface. In this study, the magnetic synoptic charts with spatial resolution of $360^\circ \times
180^\circ$ are used. To get the best extrapolation results, the order of the harmonic coefficients is chosen to be 125.  Once the coronal magnetic field is extrapolated, the average gradient of the magnetic energy density can be easily calculated for any CMEs of interest.

\section{Observations and model analyses of four cases}\label{sec:03}
Before we show the statistical results of ten CMEs, in this section four different types of CMEs are selected to investigate in detail the deflection behaviors and their relationship with the gradient of the coronal magnetic energy density. The first case is the 12 December 2008 event (CME-1), which deflected in the latitudinal direction. The second case is the 9 April 2008 event (CME-2), which deflected in the longitudinal direction. The third case is a CME erupting on 16 November 2007 (CME-3), in which a deflection in both latitudinal and longitudinal directions was obvious. The last case is the 3 November 2008 event (CME-4), which did not show an evident deflection.

\subsection{The 12 December 2008 Event (CME-1)}
This CME first appeared in the field of view (FOV) of the STA/COR1 and the STB/COR1 at about 05:35 UT on 12 December 2008. To fit the CME with the GCS model, it is required that the CME almost fully appeared in the FOV of the coronagraph. Thus the first and the last COR1 image pairs we selected are taken at 05:35 UT and 07:35 UT, respectively, during which there are 13 image pairs (or data points) with a cadence of 10 min. Similarly, in COR2 FOV, the first and last image pairs are taken at 09:52 UT and 14:52 UT, respectively, and there are 11 image pairs with a cadence of 30 min.

This CME was associated with a filament, which erupted at about 04:00 UT on 12 December 2008. The square area in Figure \ref{fg_01}(a)  represents the source region of the CME by combining the STEREO\_A/EUVI and SOHO/MDI observations. The MDI magnetogram is superposed on the EUVI 171 \AA\ image as indicated by the red (positive) and green (negative) contours, and the EUVI image is rotated to match the time and the vantage point of the MDI data. The black curve in the square displays the filament. This filament extended over a long and narrow region, from about $77^\circ$  to $106^\circ$  in longitude and about $29^\circ$  to $49^\circ$  in latitude under the Carrington coordinate system.

Figure \ref{fg_01}(b) shows the zoomed-in image of the CME source region. As has been mentioned in the last section, the tilt angle and the half angle are fixed to a certain value for the whole evolution process of the CME. After applying a trial and error method, we find that, with the value of  $-15^\circ$ for tilt angle and $14^\circ$ for half angle, the GCS model can reach the best fitting result by a visual judgement.

\begin{table}[tb]%\vskip -100pt
%\linespread{1.5}
\begin{center}
%\tabcolsep 2pt
%\footnotesize %\centering
\caption{The fitted free parameters of the 12 December 2008 CME
derived by the GCS model with the tilt angle of
$-15^\circ$ and the half angle of $14^\circ$.}\label{tb_01}
\begin{tabular}{cccccc}
\hline
Time & $\phi_c$ &$\phi_s$ &$\theta$ \\
$[UT]$ & $[deg]$ & $[deg]$ & $[deg]$
& \raisebox{1.6ex}[0pt]{$h_f/R_s$} & \raisebox{1.6ex}[0pt]{$\kappa$} \\
\hline \multicolumn{6}{c}{COR1} \\
05:35 & 72.8 & 2.0 & 30.7 & 2.46 & 0.22 \\
05:45 & 73.6 & 2.9 & 29.3 & 2.52 & 0.23 \\
05:55 & 73.8 & 3.2 & 28.1 & 2.59 & 0.23 \\
06:05 & 72.2 & 1.7 & 28.0 & 2.66 & 0.23 \\
06:15 & 74.8 & 4.4 & 27.7 & 2.76 & 0.23 \\
06:25 & 74.5 & 4.1 & 26.6 & 2.81 & 0.24 \\
06:35 & 74.2 & 4.0 & 26.5 & 2.89 & 0.24 \\
06:45 & 75.1 & 4.9 & 25.5 & 2.98 & 0.24 \\
06:55 & 77.5 & 7.4 & 22.9 & 3.24 & 0.26 \\
07:05 & 78.7 & 8.7 & 22.9 & 3.29 & 0.26 \\
07:15 & 75.7 & 5.8 & 22.6 & 3.50 & 0.27 \\
07:25 & 76.8 & 7.0 & 20.8 & 3.67 & 0.28 \\
07:35 & 76.6 & 6.9 & 19.2 & 3.81 & 0.28 \\
\multicolumn{6}{c}{COR2} \\
09:52 & 76.4 & 8.0 & 12.2 & 7.25 & 0.29 \\
10:22 & 76.2 & 8.0 & 11.7 & 7.92 & 0.29 \\
10:52 & 74.9 & 7.0 & 11.7 & 9.17 & 0.29 \\
11:22 & 75.6 & 8.0 & 11.5 & 9.93 & 0.29 \\
11:52 & 73.3 & 6.0 & 10.9 & 11.13 & 0.29 \\
12:22 & 74.7 & 7.6 & 10.7 & 11.68 & 0.29 \\
12:52 & 74.9 & 8.1 & 9.8 &   13.02 & 0.29 \\
13:22 & 77.2 & 10.7 & 9.5 & 14.12 & 0.29 \\
13:52 & 75.0 & 8.7 & 9.5 & 14.70 & 0.29 \\
14:22 & 75.3 & 9.3 & 9.6 & 16.26 & 0.29 \\
14:52 & 74.3 & 8.6 & 9.6 & 17.36 & 0.29 \\
\hline
\end{tabular}
\end{center}
\end{table}

\begin{figure}[tb]
\centering
  \includegraphics[width=0.9\hsize]{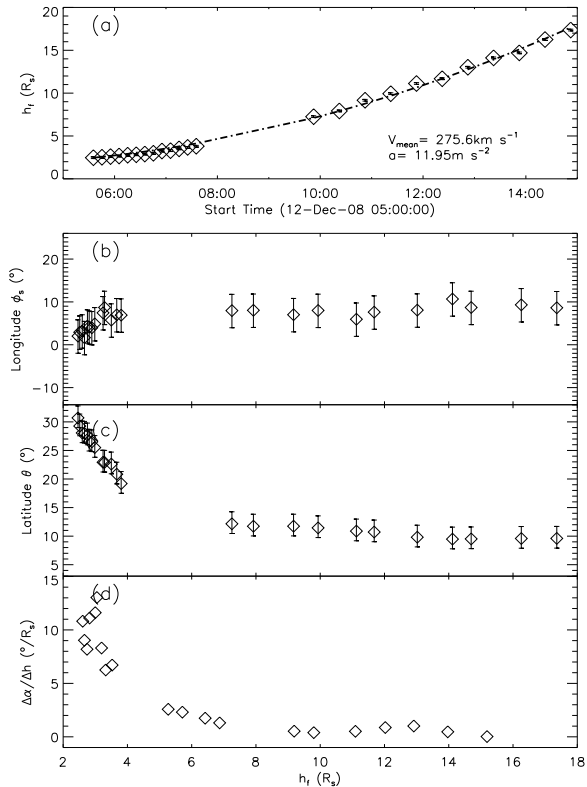}
  \caption{The kinetic evolution of the CME-1 event. Panels (a) to (d) show the height-time, Stonyhurst longitude-height, latitude-height, and deflection rate-height plots, respectively. The error bars in the first three panels are got from the 10\% decrease of the merit function (see \citet{Thernisien_etal_2009} for details).}\label{fg_03}
\end{figure}

Further, we fit the observed CME shape with the GCS model for each image pair. Figure \ref{fg_02} shows the sample of the fitting result of the CME recorded at 07:05 UT. Figures \ref{fg_02}(b1) and \ref{fg_02}(b2) present the wireframe of the GCS flux rope which is overlaid on the original images. From these two images, we found that the shapes of the CME are both consistent with the wireframe. Figures \ref{fg_02}(c1) and \ref{fg_02}(c2) present the relative brightness images derived from the GCS model. By comparing Figures \ref{fg_02}(c) with \ref{fg_02}(a), we find that they are quite similar. The arrows marked in these panels denote some common points: the bright features in Figure \ref{fg_02}(a) are also bright in Figure \ref{fg_02}(c) (arrows 1 and 3), and the darker features in Figure \ref{fg_02}(a) also look darker in Figure \ref{fg_02}(c) (arrows 2 and 4). These results indicate that the GCS model not only fits the projected shape of the CME well, but also could explain the relative brightness of the CME. Based on the above analysis, we are quite confident that the derived parameters should well reflect the 3-D geometry of the CME.

After all the 24 image pairs are processed, the CME trajectory is
obtained. All the fitted free parameters are listed in Table \ref{tb_01}. The
first column gives the time when the CME was recorded by the
STEREO/SECCHI instrument. The next two columns give the longitude
under the Carrington coordinate system (`$\phi_c$') and the
Stonyhurst coordinate system (`$\phi_s$') \citep{Thompson_2006},
respectively. The next three columns give the other geometric
parameters: latitude `$\theta$', height `$h_f$', ratio `$\kappa$'.

Figure \ref{fg_03}(a) shows the height-time plot of the CME. Both linear and
quadratic fittings are applied to the measurements. It is found that
this CME was propagating outward with a speed of about $275.6\;
km\;s^{-1}$ and an acceleration of about $12\;m\;s^{-2}$.
Figure \ref{fg_03}(b) and \ref{fg_03}(c) present the Stonyhurst longitude and latitude
as a function of the height, respectively. The Stonyhurst longitude
changed around the value of $5^\circ$ with $\approx
4^\circ$ variation. Considering the error in our fitting
process, this CME did not manifest an evident deflection in the
longitude. But its latitude shows a clear variation from about
$30^\circ$ to $10^\circ$, which suggests that
the CME experienced an evident deflection from high latitude to low
latitude.

Further, we define the deflection rate as $\triangle\alpha/\triangle
{}h$, where $\triangle\alpha$ is the deflection angle (both
latitudinal and longitudinal deflection are taken into account). For
events with more than 10 data points, the deflection rate at any
data point is calculated by fitting the longitude and the latitude
with height of neighboring five data points. For events with less
than 10 data points, a fitting procedure over three neighboring data
points is used. The variation of the deflection rate with the height
is shown in Figure \ref{fg_03}(d). It is clear that the deflection rate
decreases quickly as the height increases. The main deflection of
the CME occurred in the range below about 8 $R_s$. When the CME's
leading edge exceeded 8 $R_s$, the deflection became insignificant.
This event has been previously studied by some other researchers
\citep[e.g.][]{Byrne_etal_2010,Davis_etal_2009,Liu_etal_2010b,Poomvises_etal_2010,Lugaz_etal_2010}. The results we obtained
here are consistent with their results.

\begin{figure*}[p]
  \centering
  \includegraphics[width=0.95\hsize]{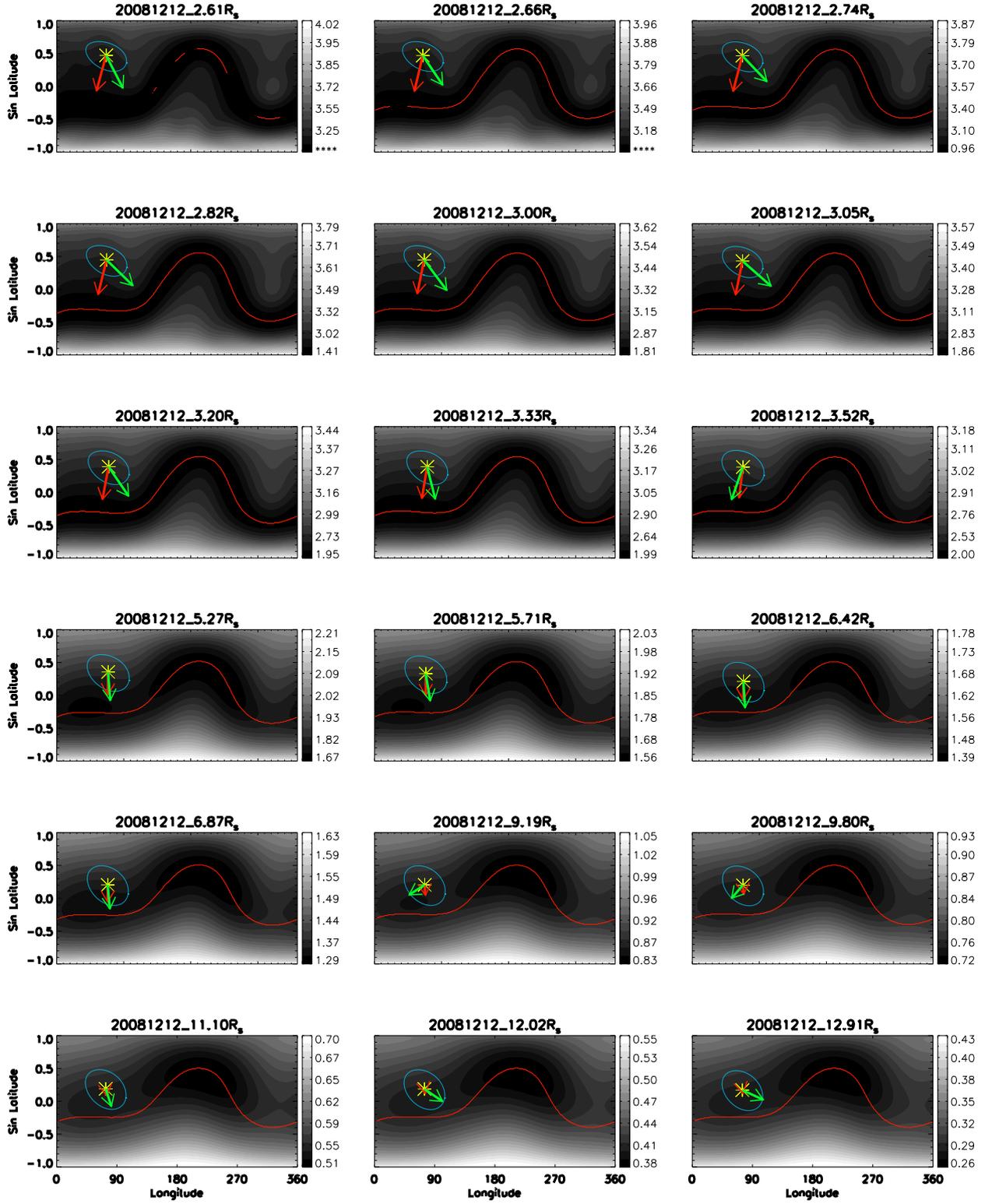}
  \caption{The comparison between the gradient of the magnetic energy density and the deflection of CME-1. The magnetic field energy density in gray scale in each panel is calculated based on the extrapolated coronal magnetic field at the corresponding altitude. The unit of the color bar of is J km$^{-3}$ in logarithm. The projected leading edge of the CME is indicated on the Carrington map by the yellow asterisk. The deflection and the gradient are represented by the green and red arrows, respectively. The lengths of the green and red arrows indicate the deflection rate and the relative strength of the gradient, respectively. The red curves indicate the heliospheric current sheet, which appears above about 2.6 $R_s$ where all the coronal magnetic field lines open.}\label{fg_04}
\end{figure*}

On the other hand, the magnetic field energy density distribution at corresponding altitude for every data points is calculated. The SOHO/MDI magnetic synoptic chart of the 2077 Carrington rotation which begins at 07:00 UT, 20 November 2008 and ends at 14:38 UT, 17 December 2008 is used as the bottom boundary for the CSSS model. Figure \ref{fg_04} shows the distributions of the magnetic energy density at different altitudes. The red curves indicate the position of the heliospheric current sheet (HCS, only marked in the panels with the altitude larger than 2.6 $R_s$, where the coronal magnetic field is open). The yellow asterisk marks the projected location of the CME leading edge on the Carrington map, and the cyan ellipse indicates the boundary of the CME in projection. The average value of the gradient of the magnetic energy density in the ellipse is marked by the red arrow, and the corresponding CME deflection is marked by the green arrow. The length of the arrows indicates the relative strength of the gradient and the deflection rate. The length is scaled by comparing its strength with all the data points of the ten CMEs. From the figure, it can be seen that the CME deflection is consistent well with the gradient of the magnetic energy density in both strength and direction, which roughly points from high latitude to low latitude. As a consequence, the CME leading edge was getting closer to the HCS during its propagation.

\subsection{The 9 April 2008 Event (CME-2)}
This CME first appeared in the FOVs of the STA/COR1 and STB/COR1 at about 10:45 UT on 9 April 2008. To guarantee that the CME almost fully appeared in the FOV, the first and last image pairs of COR1 data were taken at 10:45 UT and 11:25 UT, respectively. There are five image pairs during the interval. The first and last images of COR2 data are taken at 13:22 UT and 14:52 UT, respectively, and a total of four image pairs are selected.

\begin{figure*}[tbph]
  \centering
  \includegraphics[width=0.95\hsize]{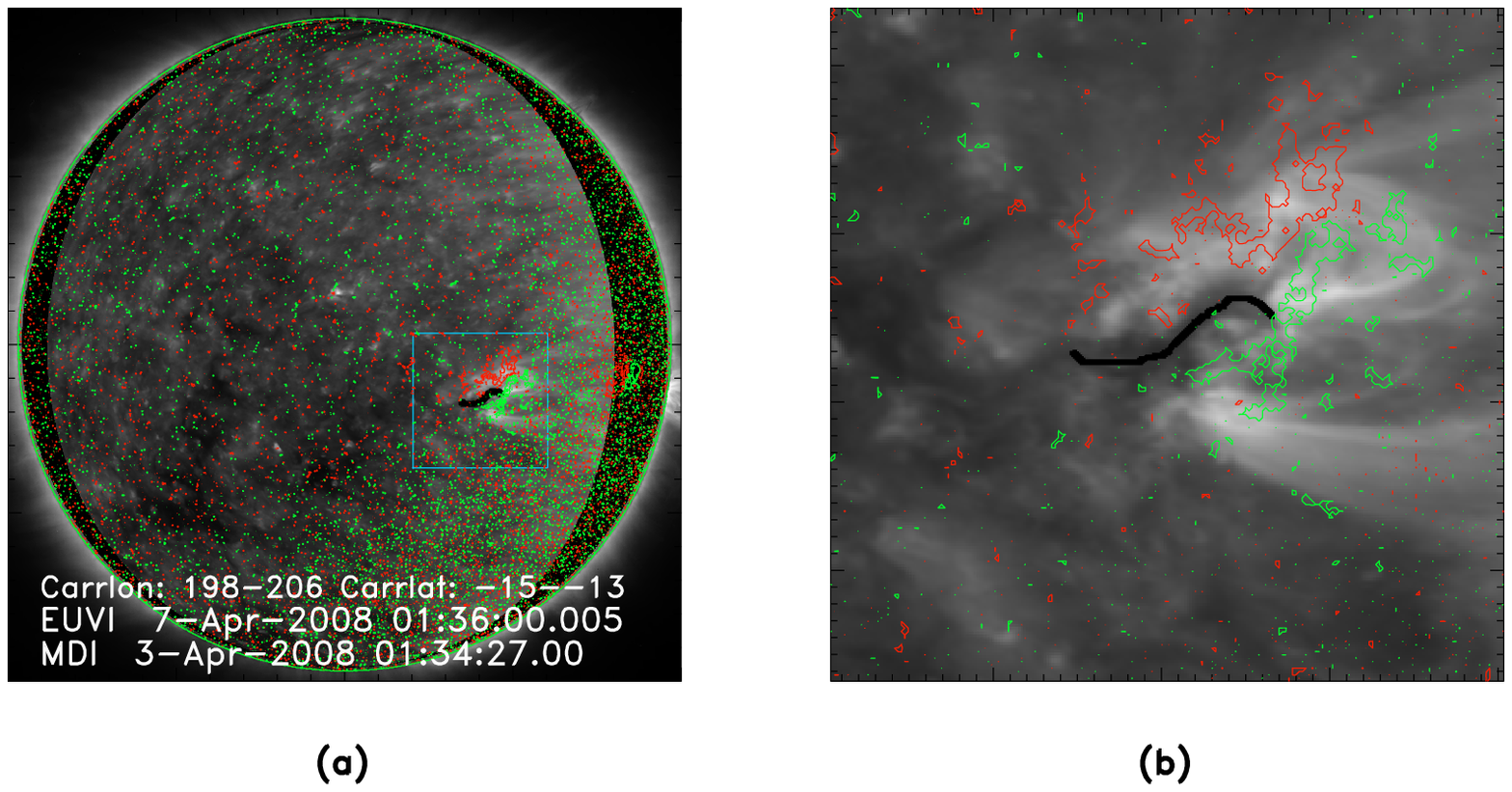}
  \caption{The source region of the 9 April 2008 CME observed by the STEREO/EUVI and the SOHO/MDI. Panel (a) shows the STEREO/EUVI 171 \AA\ image superposed by the contours of the SOHO/MDI magnetogram taken two days before the event when the source region (marked by the square box) was visible to the SOHO. The STEREO/EUVI image was rotated to match the angle of view of the SOHO/MDI. Panel (b) shows the zoomed-in image of the source region.}\label{fg_05}
%\end{figure*}
%\begin{figure*}[b]
  \centering
  \includegraphics[width=0.95\hsize]{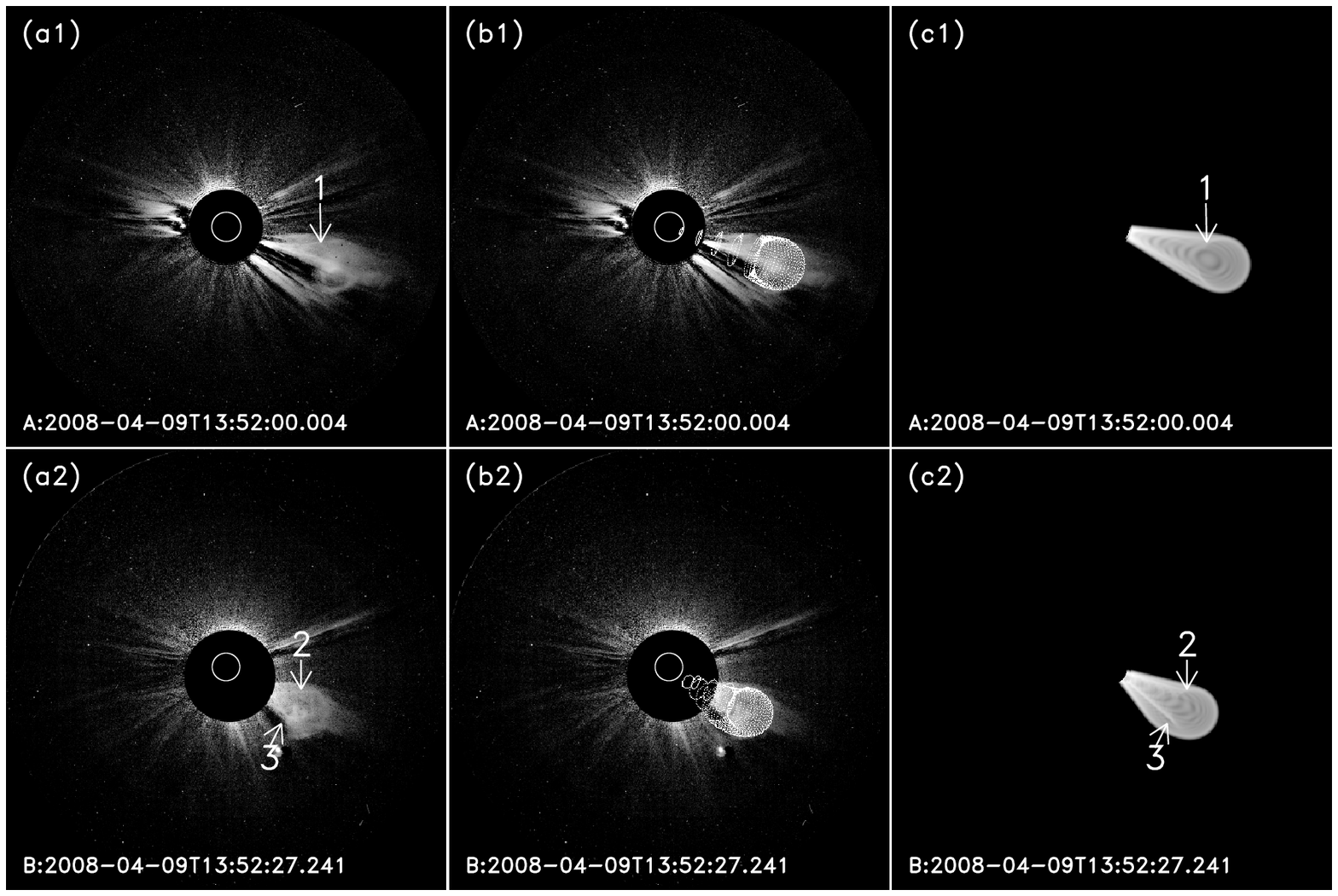}
  \caption{The fitting example of CME-2. From (a) to (c): The original CME images, the modeled wireframe images which overlays on the CME images, and the relative brightness derived from the GCS model. The top and bottom panels present the results based on the STA and STB data, respectively. The arrows with numbers indicate the common points of the CME between the original images and the derived brightness images.}\label{fg_06}
\end{figure*}

The CME was associated with an eruptive filament which erupted at
about 09:21 UT on 9 April 2008 seen from the STB. Figure \ref{fg_05}(a)
represents the combined image of the STEREO\_A/EUVI 171 \AA\ data
and the SOHO/MDI data. The square box denotes the source region of
the CME. The filament marked by the black line located from about
$198^\circ$ to $206^\circ$ in Carrington
longitude and about $-13^\circ$ to $-15^\circ$
in latitude. Figure \ref{fg_05}(b) shows the detailed image of the source
region.

\begin{table}[tbh]%\vskip -100pt
\begin{center}
%\tabcolsep 2pt
%\footnotesize %\centering
\caption{The fitted free parameters which derived by the model with the tilt angle of $8^\circ$ and the half angle of $11^\circ$ of the 9 April 2008 CME.}\label{tb_02}
\begin{tabular}{cccccc}
\hline
Time & $\phi_c$ &$\phi_s$ &$\theta$ \\
$[UT]$ & $[deg]$ & $[deg]$ & $[deg]$
& \raisebox{1.6ex}[0pt]{$h_f/R_s$} & \raisebox{1.6ex}[0pt]{$\kappa$} \\
\hline \multicolumn{6}{c}{COR1} \\
10:45 & 187.6 & 96.6 & -21.9 & 2.29 & 0.22 \\
10:55 & 190.1 & 99.2 & -21.2 & 2.53 & 0.22 \\
11:05 & 192.5 & 101.7 & -21.1 & 2.75 & 0.22 \\
11:15 & 193.5 & 102.9 & -20.1 & 3.03 & 0.22 \\
11:25 & 193.3 & 102.8 & -19.3 & 3.30 & 0.22 \\
\multicolumn{6}{c}{COR2} \\
13:22 & 197.8 & 108.2 & -18.6 & 8.50 & 0.22 \\
13:52 & 198.8 & 109.5 & -18.9 & 9.85 & 0.22 \\
14:22 & 201.3 & 112.3 & -19.1 & 11.46 & 0.22 \\
14:52 & 201.6 & 112.9 & -18.6 & 12.70 & 0.22 \\
\hline
\end{tabular}
\end{center}
\end{table}

Similar to the fitting procedure applied to CME-1, we fix the tilt angle and half angle to $8^\circ$ and $11^\circ$, respectively, by trial and error. Then we fit the CME shapes for each image pair with the GCS model. Figure \ref{fg_06} shows the sample of the fitting result of the CME which was recorded at 13:52 UT. The wireframe of the model matches well with the CME shapes viewed in both STA and STB spacecraft (Figures \ref{fg_06}(a) and \ref{fg_06}(b)). Figures \ref{fg_06}(c1) and \ref{fg_06}(c2) present the relative brightness of the CME derived from the GCS model. They are quite similar with the observed bright structure in Figures \ref{fg_06}(a1) and \ref{fg_06}(a2). The arrows in the figure mark some example common points between the modeled relative brightness images and the observed images. Thus, we believe that the 3-D geometry of the CME is reproduced by the GCS model. Table \ref{tb_02} lists the other parameters derived by the model with the tilt angle of $8^\circ$ and the half angle of $11^\circ$ of all the 9 image pairs of the CME.

\begin{figure}[tbh]
\centering
  \includegraphics[width=0.9\hsize]{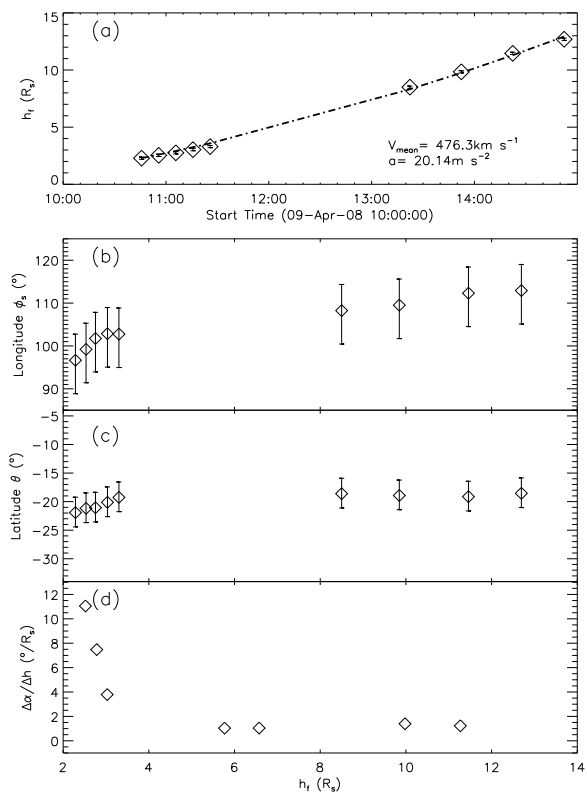}
  \caption{The kinetic evolution of the CME-2 event. Panels (a) to (d) show the height-time, longitude-height, latitude-height, and deflection rate-height curves, respectively. The error bars in the first three panels are obtained from the 10\% decrease of the merit function (see \citet{Thernisien_etal_2009} for details).}\label{fg_07}
\end{figure}

Figure \ref{fg_07}(a) shows the height-time plot of the CME. The average speed of the CME is 476.3 km s$^{-1}$, and the average acceleration is 20 m s$^{-2}$. The variations of the Stonyhurst longitude and latitude of the CME are shown in Figures \ref{fg_07}(b) and \ref{fg_07}(c). Different from CME-1, this CME manifested a weak deflection in the longitudinal direction, but no obvious deflection in the latitudinal direction. Its longitude systematically changed by about $16^\circ$ from  $\approx97^\circ$ to  $\approx113^\circ$ though the errors are large. The deflection rate of the CME is presented in Figure \ref{fg_07}(d). For CME-2, the fitting of the longitude and latitude with height over three neighboring data points is used to calculated the deflection rate as there are just nine data points in total. Similar to CME-1, the deflection mainly occurred at the low altitude where the deflection rate is as large as  $\approx10^\circ /R_s$, and it quickly decreased to about $1^\circ /R_s$ beyond  $\approx$3-5 $R_s$.

\begin{figure*}[tbph]
  \centering
  \includegraphics[width=\hsize]{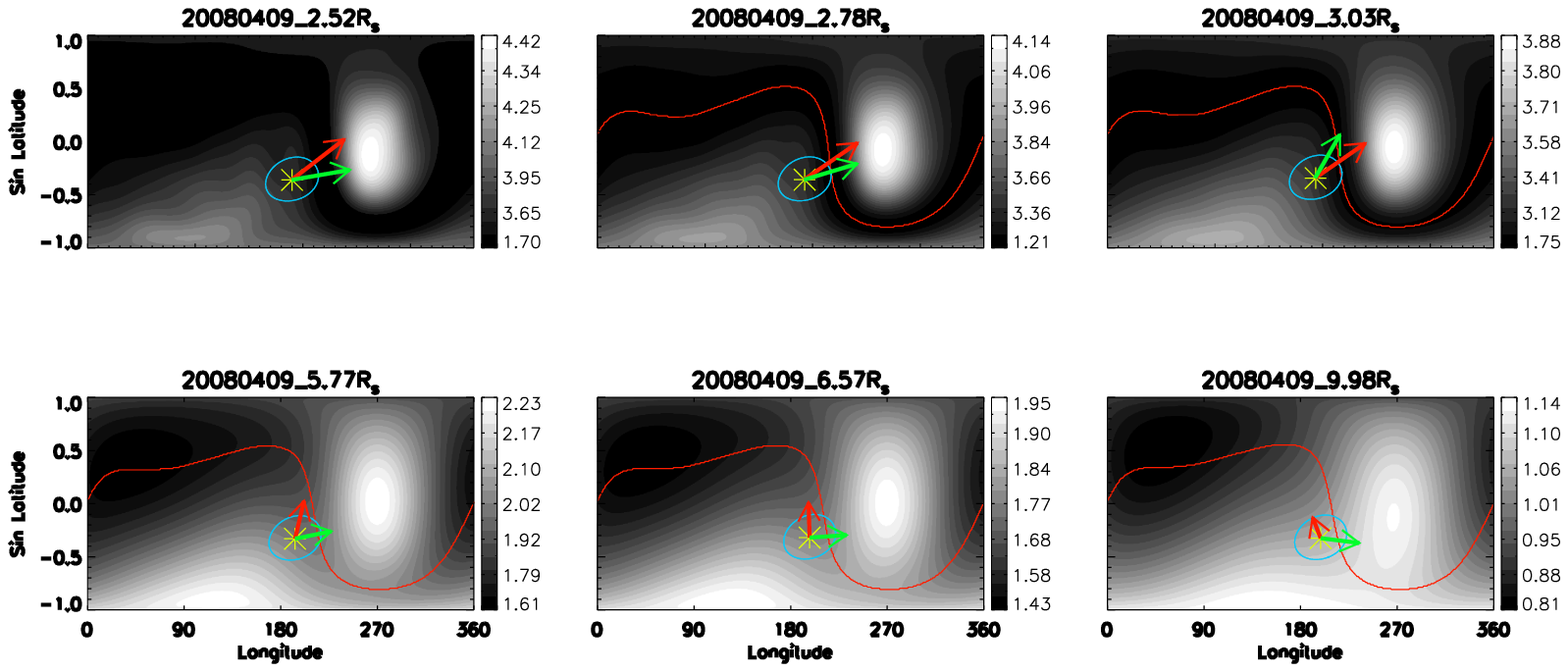}
  \caption{Same as the Figure \ref{fg_04}, but for the 9 April 2008 CME.}\label{fg_08}
%\end{figure*}
%\begin{figure*}[tbh]
  \centering
  \includegraphics[width=\hsize]{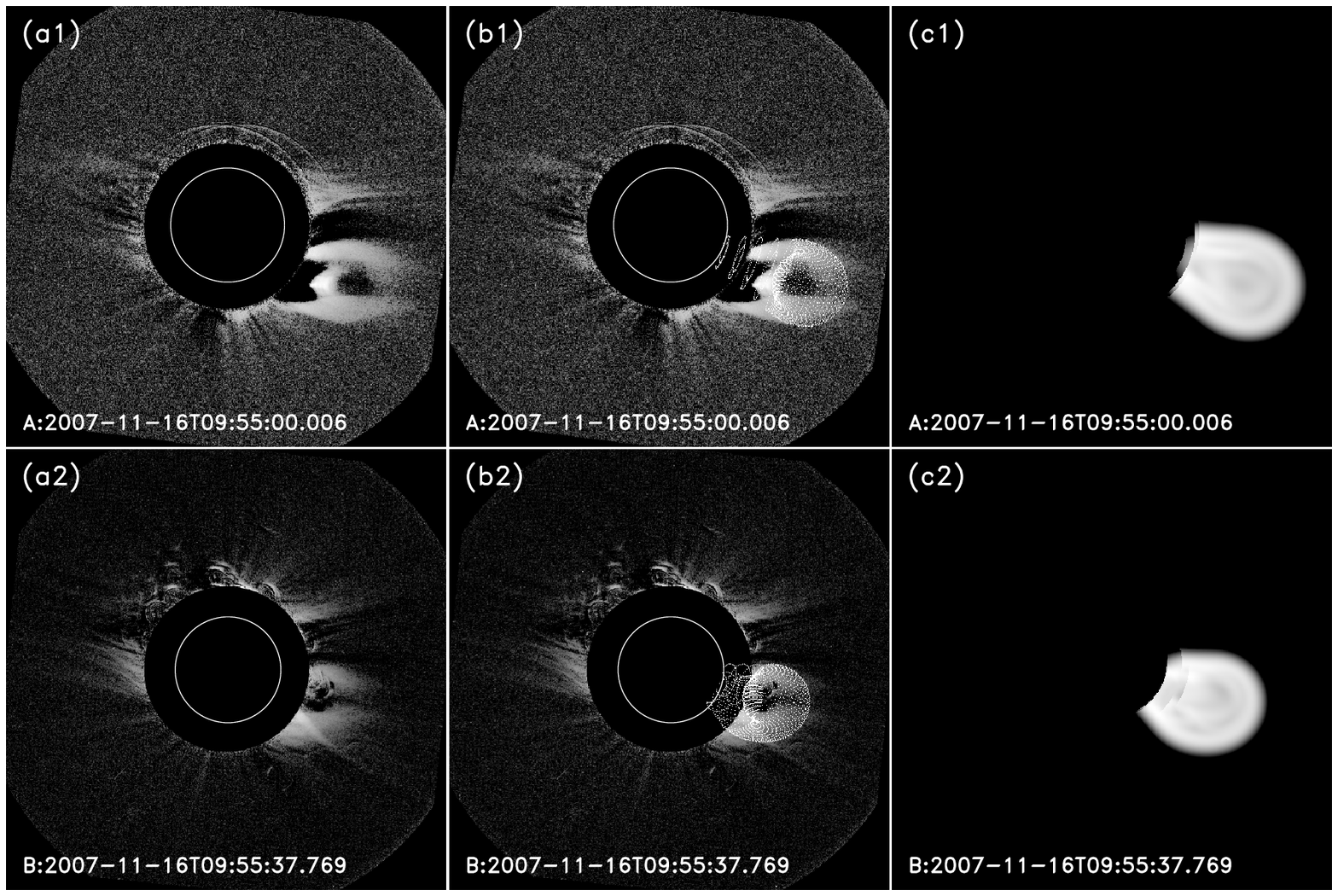}
  \caption{The fitting example of CME-3. From (a) to (c): The original CME images, the modeled wireframe images which overlays on the CME images, and the relative brightness derived from the GCS model. The top and bottom panels present the results based on the STA and STB data, respectively.}\label{fg_09}
\end{figure*}

The deflections of all the data points and the magnetic field energy density distributions of CME-2 are compared as shown in Figure \ref{fg_08}. The synoptic chart of the 2068 Carrington rotation, which begins at 01:18 UT, 20 March 2008 and ends at 08:09 UT, 16 April 2008 is used as the bottom boundary of the CSSS model. From the figure, it can be seen that the gradient direction of the magnetic energy density is mainly aligned to the deflection at a low altitude. While at a higher altitude, the angle between the two arrows becomes bigger. However, the gradient of the magnetic energy density and the deflection rate both decreased to quite low levels at the high altitude. Same as CME-1, the CME leading edge also approached close to the HCS during the propagation.

\subsection{The 16 November 2007 Event (CME-3)}
CME-3 first appeared in the COR1 FOV at about 07:35 UT on 16 November 2007. The first and last image pairs of the COR1 data were selected at 09:35 UT and 10:35 UT, respectively. During the interval, there are seven data points. The first and the last images of the COR2 data are taken at 13:52 UT and 15:22 UT, respectively, and a total of four data points are selected.

\begin{figure*}[tbhp]
  \centering
  \includegraphics[width=\hsize]{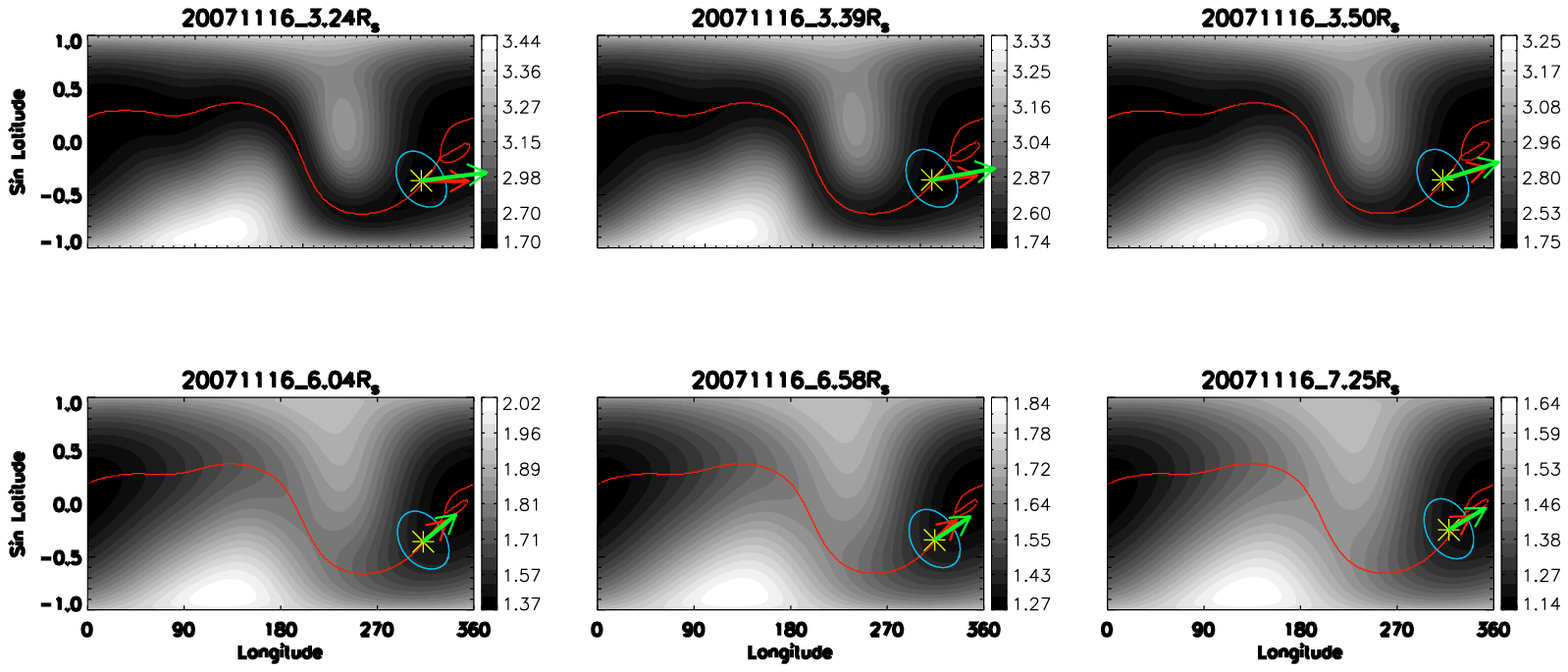}
  \caption{ Same as the Figure \ref{fg_04}, but for the 16 November 2007 event.}\label{fg_11}
%\end{figure*}
%\begin{figure*}[tbh]
  \centering
  \includegraphics[width=\hsize]{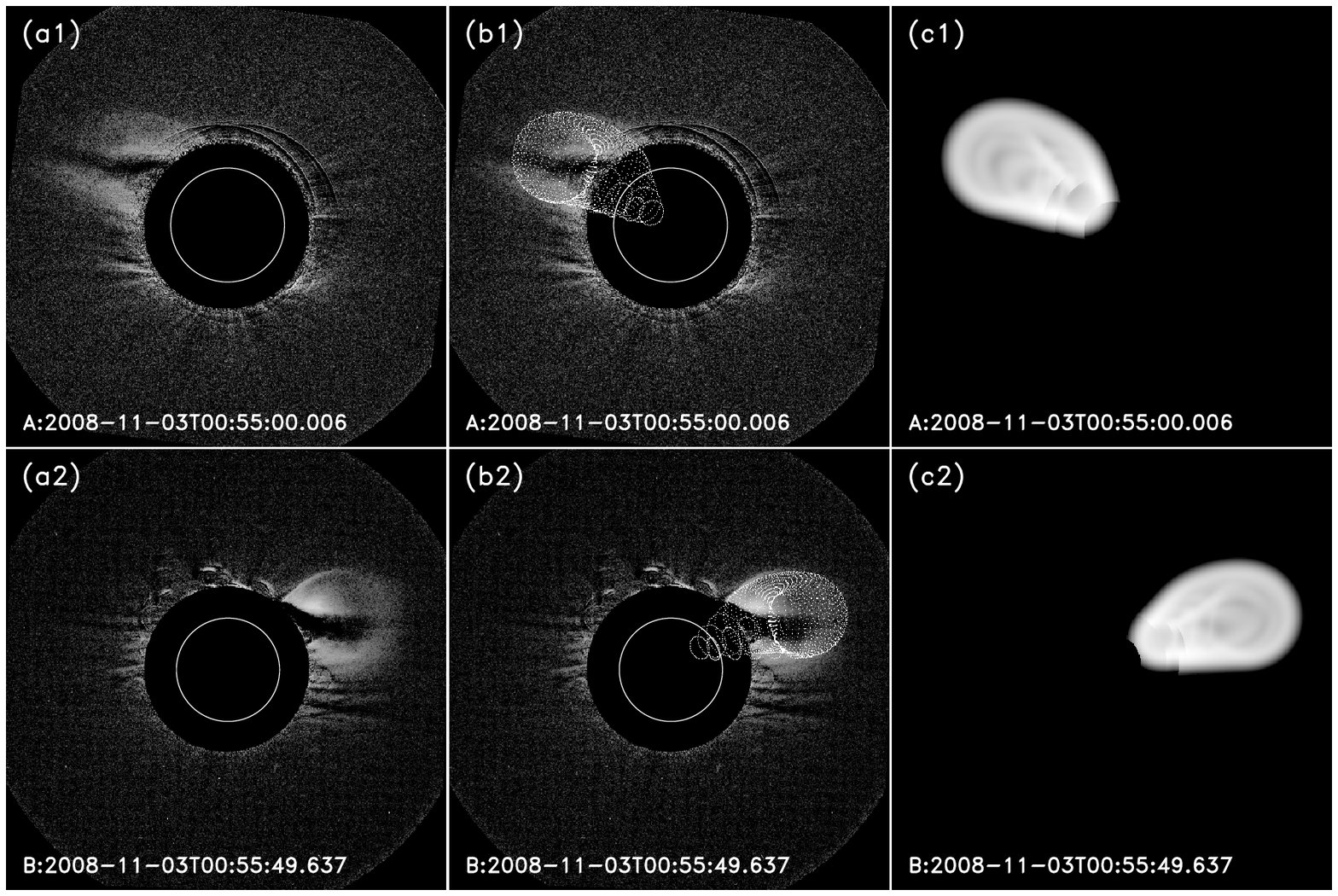}
  \caption{The fitting example of CME-4. From (a) to (c): The original CME images, the modeled wireframe images which overlays on the CME images, and the relative brightness derived from the GCS model. The top and bottom panels present the results based on the STA and STB data, respectively.}\label{fg_12}
\end{figure*}

Although CME-3 erupted from the front side of the solar surface, no significant surface activity was found. Therefore, all the parameters were obtained by the image fitting. For this event, the optimized tilt angle and half angle are $-25^\circ$ and $8^\circ$, respectively, when the GCS model can reach the best fitting of the CME images. As shown in Figure \ref{fg_12}, the GCS flux rope can fit the CME fairly well. Table \ref{tb_03} lists the parameters of the event.

\begin{table}[tbh]%\vskip -100pt
\begin{center}
%\tabcolsep 2pt
%\footnotesize %\centering
\caption{The fitted free parameters which derived by the model with the tilt angle of $-25^\circ$ and the half angle of $8^\circ$ of the 16 November 2007 CME.}\label{tb_03}
\begin{tabular}{cccccc}
\hline
Time & $\phi_c$ &$\phi_s$ &$\theta$ \\
$[UT]$ & $[deg]$ & $[deg]$ & $[deg]$
& \raisebox{1.6ex}[0pt]{$h_f/R_s$} & \raisebox{1.6ex}[0pt]{$\kappa$} \\
\hline \multicolumn{6}{c}{COR1} \\
09:35 & 303.1  & 101.1  & -22.2  & 3.04 & 0.25 \\
09:45 & 306.9  & 105.0  & -22.1  & 3.21 & 0.25 \\
09:55 & 310.2  & 108.3  & -21.6  & 3.29 & 0.25 \\
10:05 & 310.7  & 108.9  & -21.2  & 3.36 & 0.26 \\
10:15 & 311.6  & 110.0  & -21.2  & 3.43 & 0.26 \\
10:25 & 312.8  & 111.3  & -21.0  & 3.56 & 0.26 \\
10:35 & 314.2  & 112.7  & -20.0  & 3.71 & 0.26 \\
\multicolumn{6}{c}{COR2} \\
13:52 & 318.1  & 118.5  & -14.3  & 8.71 & 0.26 \\
14:22 & 321.8  & 122.4  & -14.7  & 9.73 & 0.27 \\
14:52 & 322.9  & 123.8  & -14.2  & 10.93 & 0.27 \\
15:22 & 322.4  & 123.6  & -13.6  & 11.97 & 0.27 \\
\hline
\end{tabular}
\end{center}
\end{table}

\begin{figure}[tbh]
\centering
  \includegraphics[width=0.9\hsize]{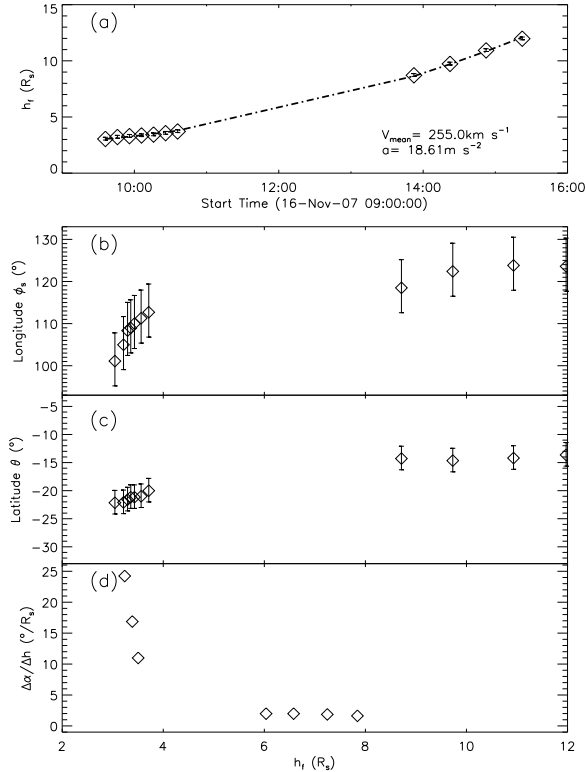}
  \caption{The kinetic evolution of the 16 November 2007 event.}\label{fg_10}
\end{figure}

Figure \ref{fg_10}(a) shows the height-time plot of the CME. The average speed of the CME is 255.0 km s$^{-1}$, and the average acceleration is 19 m s$^{-2}$. The variations of the Stonyhurst longitude and latitude of the CME are shown in Figures \ref{fg_10}(b) and \ref{fg_10}(c). Different from the above two events, the CME manifested an evident deflection in both longitude and latitude. The Stonyhurst longitude systematically changed from  $\approx101^\circ$ to  $\approx124^\circ$ while the latitude systematically changed from $\approx-22^\circ$ to $\approx-14^\circ$. The deflection rate is presented in Figure \ref{fg_10}(d). It is found that the event has higher deflection rate at the lower altitudes, and the deflected rate quickly decreased to about $1^\circ /R_s$ beyond $\approx6\;R_s$.

The deflections and the magnetic field energy density distributions of all the data points of CME-3 are compared as shown in Figure \ref{fg_11}. The synoptic chart of the 2063 Carrington rotation, which begins at 10:03 UT, 4 November 2007 and ends at 17:29 UT, 1 December 2007 is used. From this figure, it is found that the gradients of the magnetic energy density are well aligned to the deflections. The previous two CMEs occurred far away from the HCS, and then deflected towards the HCS. However, this CME almost initially originated near the HCS, and deflected along the HCS.

\subsection{The 3 November 2008 Event (CME-4)}
\begin{table}[tbh]%\vskip -100pt
\begin{center}
%\tabcolsep 2pt
%\footnotesize %\centering
\caption{The fitted free parameters derived by the model with the tilt angle of $-10^\circ$ and the half angle of $11^\circ$ of the 3 November 2008 CME.}\label{tb_04}
\begin{tabular}{cccccc}
\hline
Time & $\phi_c$ &$\phi_s$ &$\theta$ \\
$[UT]$ & $[deg]$ & $[deg]$ & $[deg]$
& \raisebox{1.6ex}[0pt]{$h_f/R_s$} & \raisebox{1.6ex}[0pt]{$\kappa$} \\
\hline \multicolumn{6}{c}{COR1} \\
00:05 & 235.5 & 7.6 & 22.6 & 3.22 & 0.23 \\
00:15 & 235.5 & 7.7 & 21.8 & 3.36 & 0.23 \\
00:25 & 234.6 & 6.9 & 21.3 & 3.61 & 0.23 \\
00:35 & 236.6 & 9.0 & 21.2 & 3.80 & 0.23 \\
00:45 & 236.5 & 8.9 & 21.6 & 3.93 & 0.23 \\
00:55 & 235.9 & 8.5 & 20.5 & 4.27 & 0.23 \\
01:05 & 236.8 & 9.5 & 19.6 & 4.57 & 0.23 \\
\multicolumn{6}{c}{COR2} \\
03:22 & 236.9 & 10.8 & 18.2 & 7.51 & 0.23 \\
03:52 & 236.1 & 10.3 & 18.4 & 8.50 & 0.23 \\
04:22 & 235.5 & 9.9 & 18.4 & 9.41 & 0.23  \\
04:52 & 236.4 & 11.2 & 18.1 & 10.14 & 0.23 \\
05:22 & 235.5 & 10.5 & 18.5 & 10.95 & 0.23 \\
05:52 & 235.2 & 10.5 & 18.6 & 11.97 & 0.23 \\
06:22 & 233.6 & 9.2 & 19.0 & 12.64 & 0.23  \\
06:52 & 234.6 & 10.4 & 19.0 & 13.09 & 0.23 \\
\hline
\end{tabular}
\end{center}
\end{table}

This CME first appeared in the COR1 FOV at about 23:35 UT on 2 November 2008. The first and last images of the COR1 data were taken at 00:05 UT and 01:05 UT on 3 November 2008, respectively, and there are seven image pairs during the interval. The first and last images of the COR2 data are taken at 03:22 UT and 06:52 UT, respectively, and a total of eight data points are selected.

Same as the CME-3 event, there is no clear source region observation. The tilt angle and half angle are fixed to $-10^\circ$ and $11^\circ$, respectively, to get the best fitting of the CME shapes observed by both the STA and STB spacecraft. The GCS flux rope can fit the CME fairly well as shown in Figure \ref{fg_12}. Table \ref{tb_04} lists the parameters of all the 15 data points of the event.

\begin{figure*}[tbph]
\centering
  \includegraphics[width=0.4\hsize]{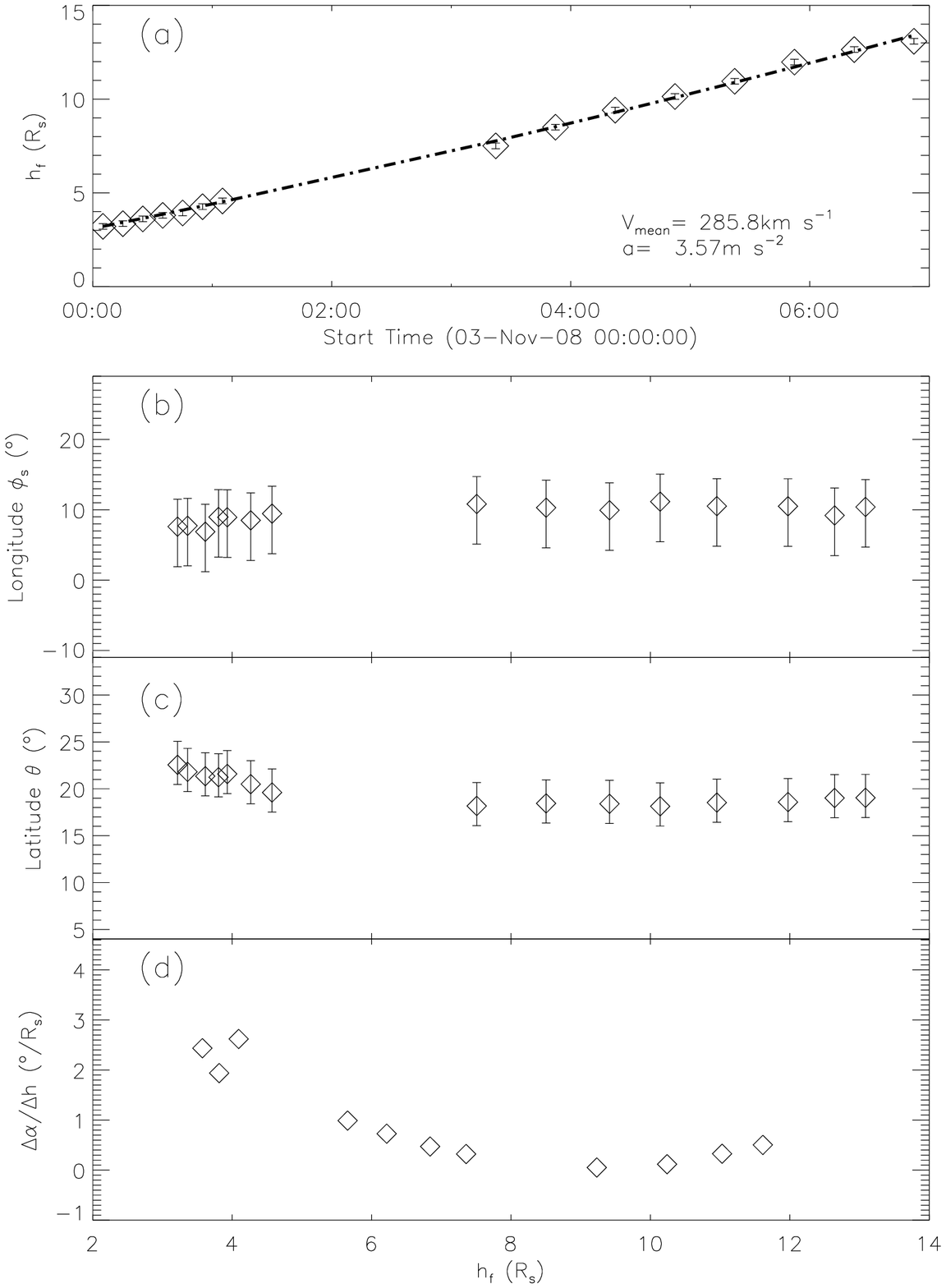}
  \caption{The kinetic evolution of the 3 November 2008 event.}\label{fg_13}
%\end{figure*}
%\begin{figure*}[tbh]
  \centering
  \includegraphics[width=\hsize]{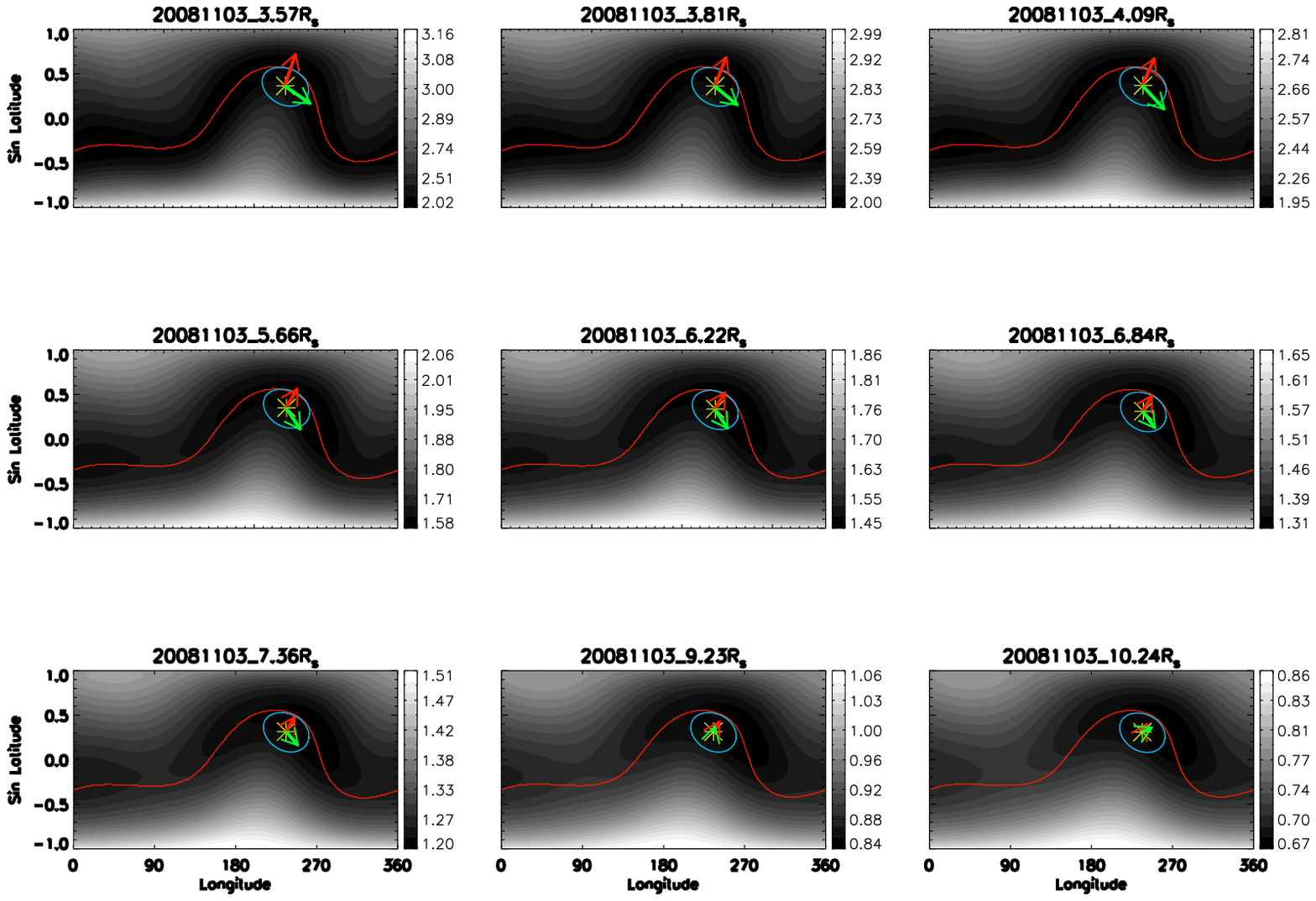}
  \caption{ Same as the Figure \ref{fg_04}, but for the 3 November 2008 event.}\label{fg_14}
\end{figure*}

Figure \ref{fg_13}(a) shows the height-time plot of the CME, and suggests a slight acceleration during the propagation. The average speed of the CME is 285.8 km s$^{-1}$, and the average acceleration is 4 m s$^{-2}$. The variations of the Stonyhurst longitude and latitude of the CME are shown in Figures \ref{fg_13}(b) and \ref{fg_13}(c), respectively. Neither the Stonyhurst longitude nor the latitude shows a significant change. There is no more than $3^\circ$ variation of the longitude. The latitude just changed slightly from $22.6^\circ$  to $19.6^\circ$  in the COR1 FOV and did not vary in the COR2 FOV. Moreover, its deflection rate is no more than $3^\circ/R_s$  as shown in Figure \ref{fg_13}(d). Thus, this CME could be taken as the event without an obvious deflection during its propagation.

\begin{table*}[b]%\vskip -100pt
\begin{center}
%\tabcolsep 2pt
%\footnotesize %\centering
\caption{The fitted free parameters of all the CME events.}\label{tb_05}
\begin{tabular}{cccccccccccc}
\hline Date & Time & $\phi_c$
&$\phi_s$ &$\theta$ & $h_f/R_s$
& $\kappa$ & $\gamma$ & $\alpha$
& $\triangle\alpha$
& $V$ & $N$\\
& $[UT]$ & $[deg]$ & $[deg]$ & $[deg]$ & & & $[deg]$ & $[deg]$ & $[deg]$ & $[km\;s^{-1}]$ \\
\hline
 & 09:35 & 303.1 & 101.1 & -22.2 & 3.0 & 0.25 \\
\raisebox{1.6ex}[0pt]{16-Nov-07} & 15:22 & 322.4 & 123.6 & -13.6 &
12.0 & 0.27 & \raisebox{1.6ex}[0pt]{-24.6} &
\raisebox{1.6ex}[0pt]{8.4} & \raisebox{1.6ex}[0pt]{24.0} &
\raisebox{1.6ex}[0pt]{255} & \raisebox{1.6ex}[0pt]{11} \\
\hline
 & 06:05 & 34.3 & 67.6 & 14.2 & 3.3 & 0.14 \\ \raisebox{1.6ex}[0pt]{04-Dec-07} &  16:52 &    34.2 & 73.4 & 3.9 & 13.8 & 0.21 & \raisebox{1.6ex}[0pt]{-56.5} & \raisebox{1.6ex}[0pt]{10.1} & \raisebox{1.6ex}[0pt]{11.9}  & \raisebox{1.6ex}[0pt]{182} & \raisebox{1.6ex}[0pt]{19} \\
\hline
 & 22:45 & 226.7 & 194.6 & -23.8 & 3.7 & 0.27 \\ \raisebox{1.6ex}[0pt]{22-Jan-08} &  06:22$^n$ & 223.9 & 196.0 & -25.2 & 16.4 & 0.36 & \raisebox{1.6ex}[0pt]{-21.8} & \raisebox{1.6ex}[0pt]{10.1} & \raisebox{1.6ex}[0pt]{1.9}  & \raisebox{1.6ex}[0pt]{296} & \raisebox{1.6ex}[0pt]{16} \\
\hline
 & 17:05 & 224.1 & 250.2 & 21.5 & 3.5 & 0.21 \\ \raisebox{1.6ex}[0pt]{23-Feb-08} &  06:22$^n$ & 211.2 & 244.6 & 19.2 & 15.5 & 0.24 & \raisebox{1.6ex}[0pt]{25.7} & \raisebox{1.6ex}[0pt]{11.7} & \raisebox{1.6ex}[0pt]{6.0}  & \raisebox{1.6ex}[0pt]{174} & \raisebox{1.6ex}[0pt]{25} \\
\hline
 & 19:05 & 194.2 & 270.0 & -12.8 & 2.5 & 0.25 \\ \raisebox{1.6ex}[0pt]{25-Mar-08} &  20:52 & 201.7 & 278.4 & -12.3 & 12.7 & 0.30 & \raisebox{1.6ex}[0pt]{35.2} & \raisebox{1.6ex}[0pt]{12.6} & \raisebox{1.6ex}[0pt]{8.5}  & \raisebox{1.6ex}[0pt]{1092} & \raisebox{1.6ex}[0pt]{4} \\
\hline
 & 16:15 & 260.8 & 120.1 & 1.3 & 3.3 & 0.17 \\ \raisebox{1.6ex}[0pt]{05-Apr-08} &  18:22 & 251.9 & 112.3 & 3.8 & 14.0 & 0.25 & \raisebox{1.6ex}[0pt]{-64.8} & \raisebox{1.6ex}[0pt]{9.8} & \raisebox{1.6ex}[0pt]{8.1}  & \raisebox{1.6ex}[0pt]{982} & \raisebox{1.6ex}[0pt]{5} \\
\hline
 & 10:45 & 187.6 & 96.6 & -21.9 & 2.3 & 0.22 \\ \raisebox{1.6ex}[0pt]{09-Apr-08} &  14:52 & 201.6 & 112.9 & -18.5 & 12.7 & 0.22 & \raisebox{1.6ex}[0pt]{8.4} & \raisebox{1.6ex}[0pt]{10.6} & \raisebox{1.6ex}[0pt]{16.6}  & \raisebox{1.6ex}[0pt]{476} & \raisebox{1.6ex}[0pt]{9} \\
 \hline
 & 00:05 & 235.5 & 7.6 & 22.6 & 3.2 & 0.23 \\ \raisebox{1.6ex}[0pt]{03-Nov-08} &  06:52 & 234.6 & 10.4 & 19.0 & 13.1 & 0.23 & \raisebox{1.6ex}[0pt]{-10.1} & \raisebox{1.6ex}[0pt]{11.2} & \raisebox{1.6ex}[0pt]{4.5}  & \raisebox{1.6ex}[0pt]{286} & \raisebox{1.6ex}[0pt]{15} \\
 \hline
 & 13:05 & 288.8 & 199.9 & -22.3 & 3.0 & 0.24 \\ \raisebox{1.6ex}[0pt]{13-Nov-08} &  21:22 & 275.5 & 191.1 & -12.1 & 16.4 & 0.27 & \raisebox{1.6ex}[0pt]{-30.2} & \raisebox{1.6ex}[0pt]{11.2} & \raisebox{1.6ex}[0pt]{13.4}  & \raisebox{1.6ex}[0pt]{256} & \raisebox{1.6ex}[0pt]{24} \\
 \hline
 & 05:35 & 72.8 & 2.0 & 30.7 & 2.5 & 0.22 \\ \raisebox{1.6ex}[0pt]{12-Dec-08} &  14:52 & 74.3 & 8.7 & 9.6 & 17.4 & 0.29 & \raisebox{1.6ex}[0pt]{-15.1} & \raisebox{1.6ex}[0pt]{14.0} & \raisebox{1.6ex}[0pt]{22.1}  & \raisebox{1.6ex}[0pt]{276} & \raisebox{1.6ex}[0pt]{24} \\
 \hline
\end{tabular}
\end{center}
* For each event, the parameters at the first and last valid times are given in
two rows. Column ``Date'' gives the date when the CME occurred. The
second column lists the time when the CME observed and the
superscript 'n' means the time of the next day. The next seven
columns give the model parameters: the Carrington longitude
`$\phi_c$', the Stonyhurst longitude `$\phi_s$', latitude
`$\theta$', height `$h_f$', ratio `$\kappa$', tilt angle `$\gamma$',
and half angle `$\alpha$'. The 10th column means the solid angle
between the first and the last data point. The next two columns give
the average speed of the event and the total number of data points.
\end{table*}

The comparison between the deflections and the magnetic field energy density distributions of all the data points are presented in Figure \ref{fg_14}. The synoptic chart of the 2076 Carrington rotation which begins at 23:42 UT, 23 October 2008 and ends at 07:00 UT, 20 November 2008 is adopted as the bottom boundary of the CSSS model. From the figure, it can be seen that the gradient direction of the magnetic energy density is toward to the nearby HCS. The changes in the CME propagation direction almost do not align with the directions of the gradient. Compared to the previous three events, both the deflection rate and the gradient for this event are quite small. Thus the large deviation between the two directions might not be an inconsistency.

\section{Statistical Analysis of CME Deflections}\label{sec:04}

In the above analysis, we present four events which manifest different deflection properties during their propagation in the corona. The first two events: the 12 December 2008 event (CME-1) and the 9 April 2008 event (CME-2), which appeared apart from the heliospheric current sheet at the early stage, deflected basically along the gradient direction of the magnetic energy density. Both of them approached toward the HCS which is generally located at the region with the lowest magnetic energy density. The 16 November 2007 event (CME-3), which initially originated near the HCS, manifested a deflection along the HCS that is in both longitudinal and latitudinal directions. The 3 November 2008 event (CME-4) did not exhibit an evident deflection, and accordingly the gradient of magnetic energy density was also very small.

\begin{figure*}[tbhp]
\centering
  \includegraphics[width=0.85\hsize]{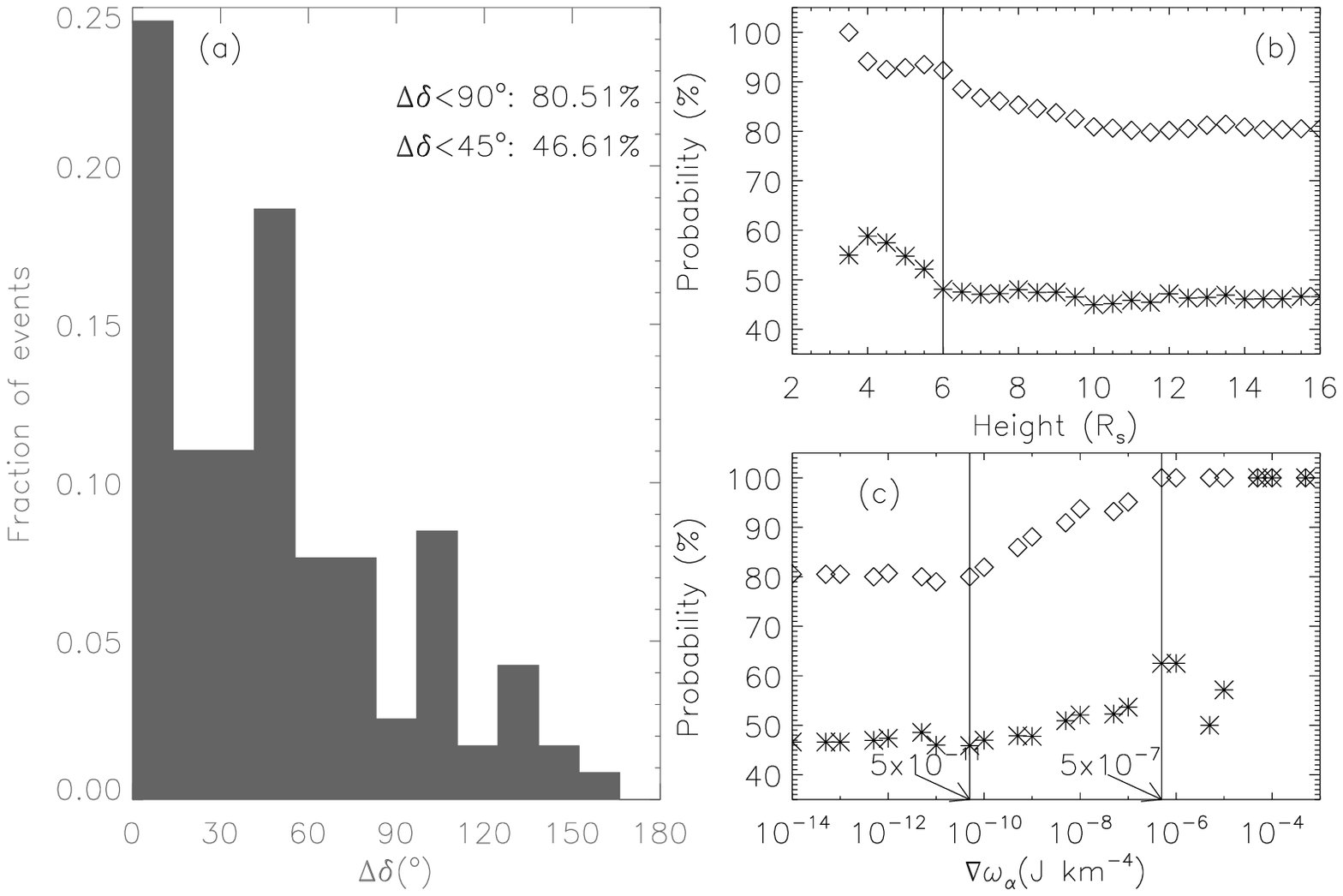}
  \caption{Statistical analysis of the angle between the deflection direction and the magnetic energy density gradient. (a) The distribution of the angle. (b) The probabilities of the angle $\leq90^\circ$(diamond) and $\leq45^\circ$ (asterisk) as a function of height. (c) Same as the figure (b), but presents the probabilities as a function of the strength of the magnetic energy density gradient.}\label{fg_15}
%\end{figure*}
%\begin{figure*}[tbh!]
\centering
  \includegraphics[width=0.85\hsize]{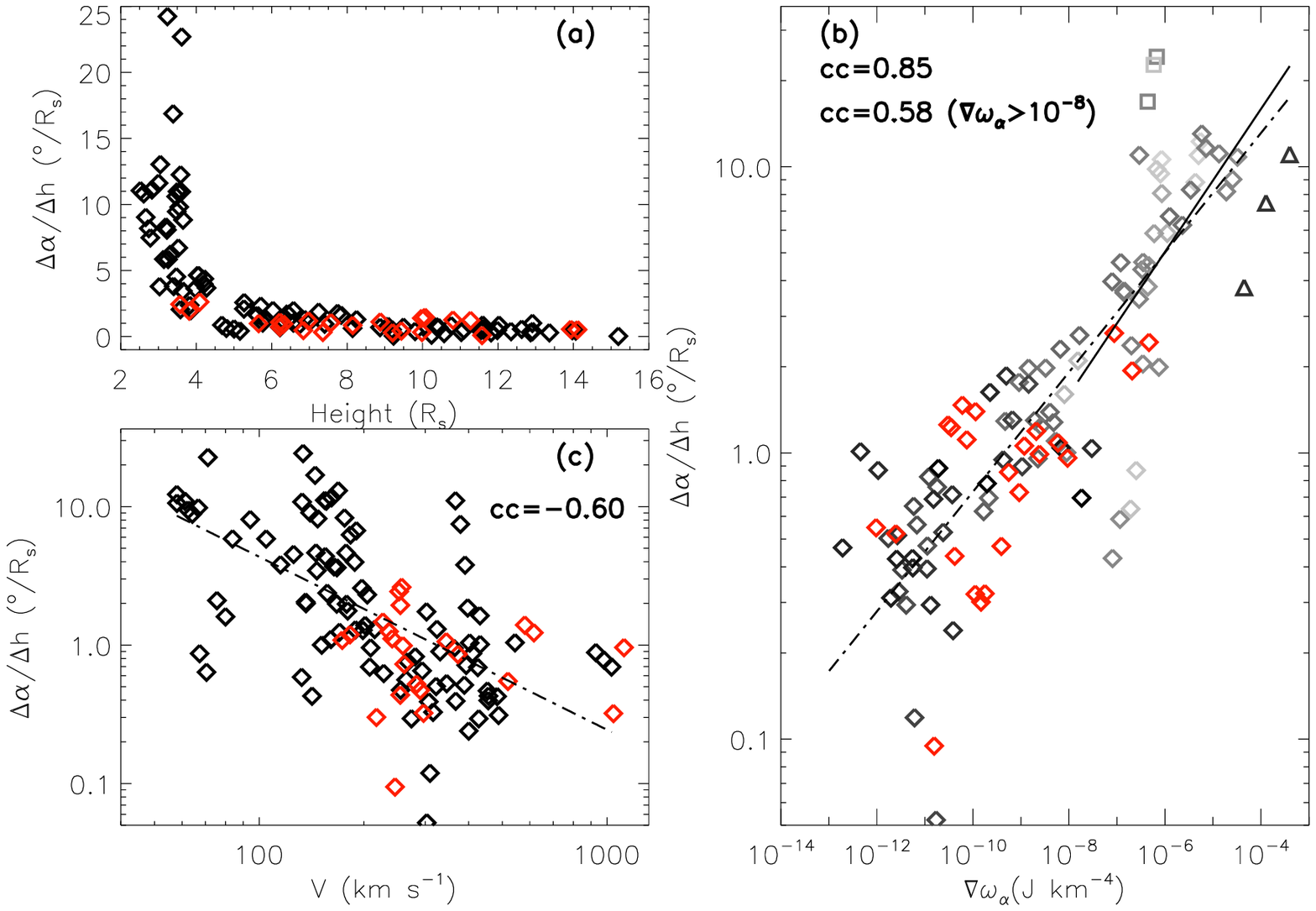}
  \caption{ Quantitative analysis about the deflection rate. The red diamonds mark the `bad' points, at which the deflection direction is opposite to the gradient direction. Three panels show the scatter plots between the deflection rate versus (a) height, (b) the strength of the gradient, and (c) the instantaneous radial speed. In panel (b) the instantaneous radial speed of a CME is coded in gray scale; darker colors stand for larger speeds. }\label{fg_16}
\end{figure*}

\begin{figure*}[tb]
\centering
  \includegraphics[width=\hsize]{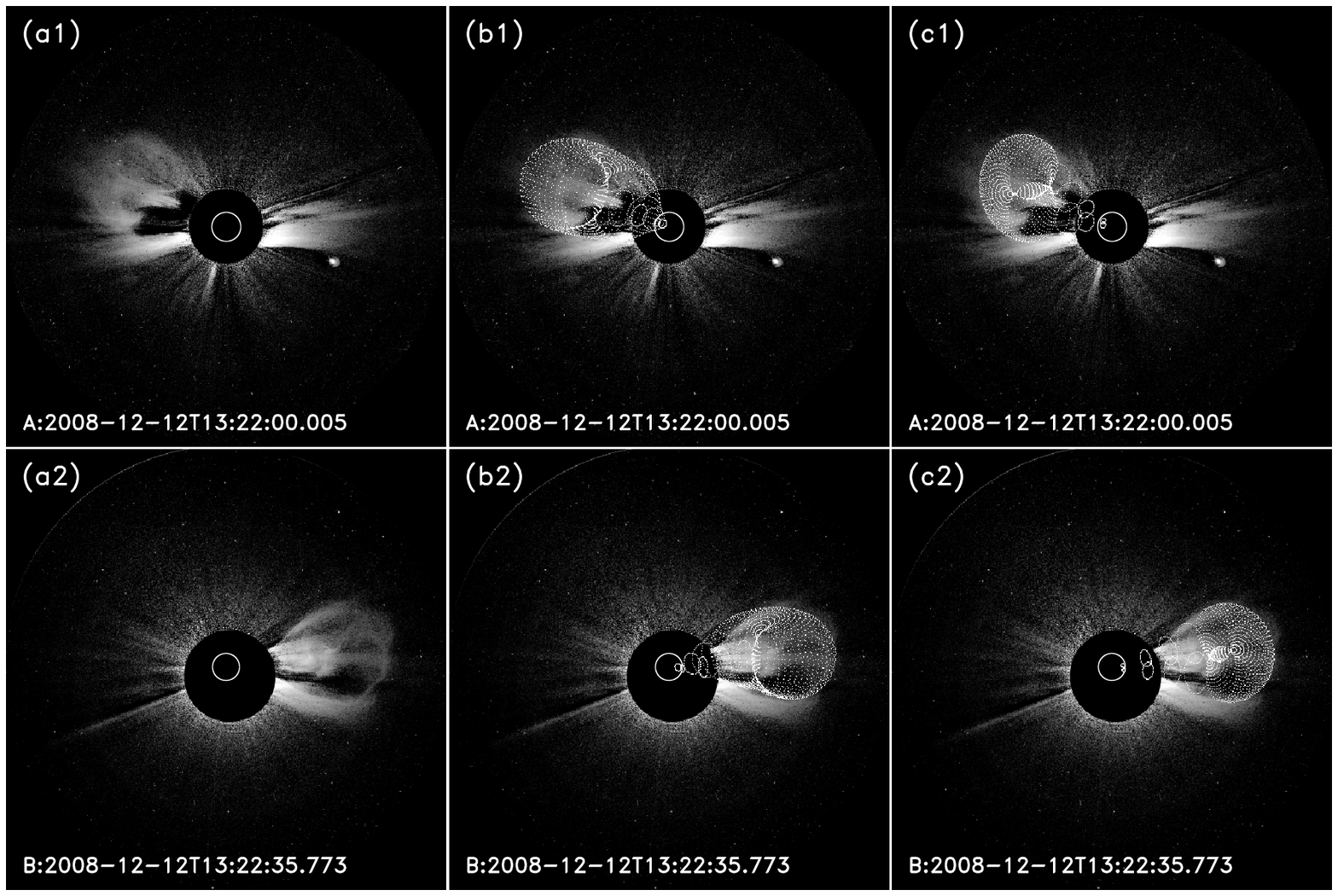}
  \caption{The fitting example of CME-1 due to the GCS model with the tilt angles of $-15^\circ$ and $-84^\circ$. From (a) to (c): The original CME images, the modeled wireframe images with the tilt angle of $-15^\circ$ and $-84^\circ$. The top and bottom panels present the results based on the STA and STB data, respectively.}\label{fg_17}
\end{figure*}

Besides the above four events, another six events which have clear observations in the FOVs of the COR1 and the COR2 during the period from November 2007 to the end of 2008 have also been studied. The main parameters of all the ten events are listed in Table \ref{tb_05}. For each event we give the parameters at the first and the last valid times and list them in two rows.

It could be found from Table \ref{tb_05} that there are two events, 22 January 2008 and 3 November 2008 (CME-4), which did not deflect obviously. The other events all manifested a deflection during the propagation. Especially the events erupting on 16 November 2007 (CME-3) and 13 November 2008 deflected in both longitudinal and latitudinal directions. Similar to CME-3, the 13 November 2008 CME also deflected along the HCS.

For all the events, we have a total of 118 data points of deflection, magnetic energy density gradient, and the corresponding information such as height, the instantaneous velocity, etc. Figure \ref{fg_15}(a) shows the distribution of the angle between the directions of the deflection and the gradient of magnetic energy density. It can be read from the histogram that the fraction of events decreases from small angles (agreement between the direction of the gradient of magnetic energy density and the deflection) to large angles (disagreement). The bin of the angle $\leq15^\circ$ has the most data points, almost half the data points have the angle $\leq45^\circ$, and as much as 80\% data points have the angle $\leq90^\circ$. The angle $\leq90^\circ$ means that the deflections are marginally consistent with the gradient directions, and the angle $\leq45^\circ$ indicates a good consistency. Figure \ref{fg_15}(b) presents the probability of the angle $\leq90^\circ$ (diamond) and $\leq45^\circ$ (asterisk), respectively, as a function of the height of the CME leading edge. At any height, there are at least 80\% of data points with the angle $\leq90^\circ$, and at least 45\% of data points with the angle $\leq45^\circ$. Particularly, the probabilities are higher within about 6 $R_s$, which suggests that the deflections and the gradients have a better consistency in the inner corona. Figure \ref{fg_15}(c) presents the probability versus the strength of the magnetic energy density gradient. It is clearly shown that the deflections and the gradients have a better consistency when the gradient is stronger. When the gradient is larger than  $5\times10^{-11}J\;km^{-4}$, the two directions are marginally consistent, while the gradient is larger than $5\times10^{-7}J\;km^{-4}$, they are highly consistent.

A further quantitative analysis about the deflection rate is shown
in the Figure \ref{fg_16}. The data points with the angle between the
directions of the deflection and the magnetic energy density gradient larger than
$90^\circ$ are defined as `bad points' and indicated by
the red diamonds. Figure \ref{fg_16}(a) shows the deflection rate as a
function of the height. It is found that the deflection rate of the
CMEs decreases quickly with increasing height. The CMEs generally
have large deflection rate within about 4 $R_s$, and in the outer
corona, the deflection rate approaches to zero. None of the bad
points is beyond the deflection rate of $3^\circ/R_s$.

The correlation between the deflection rate and the strength of the magnetic energy density gradient is shown in Figure \ref{fg_16}(b). The instantaneous radial speed of a CME is represented in gray scale, in which darker symbols stand for larger speeds. It is found that all the `bad points' (shown in red diamonds) appear in the area with the magnetic energy density gradient lower than $5\times10^{-7}J\;km^{-4}$ and the deflection rate within $3^\circ/R_s$. In such a region, any errors in our calculation may become relatively significant. Thus, as we have stated before, these `bad points' cannot be treated as an inconsistency. Without these bad points, there is an evident positive correlation between the deflection rate and the gradient strength. The correlation coefficient is about $0.85$. It suggests that a stronger gradient causes a larger deflection rate.  Considering that the deflection rate is not significant when the magnetic energy density gradient is lower than $10^{-8}J\;km^{-4}$, we also show the correlation between the deflection rate and the gradient for the data points with the gradient larger than $10^{-8}J\;km^{-4}$, which is represented by the solid line in Figure \ref{fg_16}(b). It still shows a positive correlation, though the correlation coefficient has decreased to $0.58$.

The relative low correlation coefficient is mainly due to significant scatter of the data points at the right-upper corner. We notice that there are three data points at strong gradients but having a relatively small deflection rate (marked by `$\triangle$'),
and other three data points with a large deflection but at relatively weak gradients (marked by `$\square$'). The two sets of data points are obviously against the overall correlation we obtained, and may imply that there should be other factors influencing the deflection rate of CMEs.

A possible factor is the CME radial speed. By comparing the deflection rate of these data points with the CME radial speed (derived from the height-time plot, see Figure \ref{fg_03}(a) for example), we found that the speeds of the data points `$\triangle$' are bigger than those of the data points `$\square$'. For all the other data points, Figure \ref{fg_16}(c) shows the deflection rate as a function of the CME radial speed. A weak anti-correlation is found between the deflection rate and the speed. The data points with a higher speed generally experience a slower deflection, i.e., the faster a CME moves outward, the smaller is the deflection rate.

Besides, the CME mass should be another important factor. Mass characterizes the inertia of a CME. Thus, the heavier a CME is, the smaller should be the deflection rate. However, there are only four events having available mass in the CDAW CME catalog (http://cdaw.gsfc.nasa.gov/CME\_list/).The event number is too small to derive a reliable result. Moreover, considering that there are significant errors in the mass determination \citep{Colaninno_Vourlidas_2009,Lugaz_etal_2005,Vourlidas_etal_2000,Wang_etal_2011}, the effect of CME mass on the deflection is not analyzed in this paper.

\section{Summary and Discussion}\label{sec:05}

\begin{table}[tb]%\vskip -100pt
\begin{center}
%\tabcolsep 2pt
%\footnotesize %\centering
\caption{The fitted free parameters of the 12 December 2008 CME derived by the GCS model with the tilt angle of $-84^\circ$ and the half angle of $8^\circ$}\label{tb_06}
\begin{tabular}{cccccc}
\hline
Time & $\phi_c$ &$\phi_s$ &$\theta$ \\
$[UT]$ & $[deg]$ & $[deg]$ & $[deg]$
& \raisebox{1.6ex}[0pt]{$h_f/R_s$} & \raisebox{1.6ex}[0pt]{$\kappa$} \\
\hline \multicolumn{6}{c}{COR1} \\
05:35 & 72.0 & 1.2 & 28.8 & 2.54 & 0.16 \\
05:45 & 75.3 & 4.6 & 28.2 & 2.56 & 0.16 \\
05:55 & 74.0 & 3.4 & 26.9 & 2.67 & 0.16 \\
06:05 & 74.6 & 4.0 & 26.8 & 2.73 & 0.16 \\
06:15 & 73.7 & 3.3 & 26.7 & 2.76 & 0.17 \\
06:25 & 74.3 & 3.9 & 26.0 & 2.81 & 0.17 \\
06:35 & 72.8 & 2.6 & 25.7 & 2.85 & 0.17 \\
06:45 & 73.1 & 2.9 & 24.4 & 3.11 & 0.17 \\
06:55 & 74.3 & 4.3 & 24.3 & 3.24 & 0.17 \\
07:05 & 74.6 & 4.6 & 23.4 & 3.37 & 0.17 \\
07:15 & 74.5 & 4.6 & 22.6 & 3.63 & 0.17 \\
07:25 & 74.5 & 4.7 & 21.6 & 3.85 & 0.18 \\
07:35 & 74.4 & 4.7 & 20.2 & 4.02 & 0.18 \\
\multicolumn{6}{c}{COR2} \\
09:52 & 73.4 & 4.9 & 12.3 & 6.99 & 0.20 \\
10:22 & 75.9 & 7.7 & 12.4 & 8.03 & 0.20 \\
10:52 & 72.8 & 4.9 & 11.8 & 9.04 & 0.20 \\
11:22 & 76.0 & 8.4 & 11.3 & 9.97 & 0.20 \\
11:52 & 74.0 & 6.6 & 10.4 & 11.33 & 0.20 \\
12:22 & 74.3 & 7.2 & 9.5 & 11.89 & 0.20 \\
12:52 & 74.5 & 7.7 & 9.4 & 13.28 & 0.20 \\
13:22 & 75.5 & 9.0 & 9.4 & 14.03 & 0.20 \\
13:52 & 74.9 & 8.7 & 8.3 & 15.13 & 0.20 \\
14:22 & 75.0 & 9.0 & 8.4 & 16.28 & 0.20 \\
14:52 & 73.2 & 7.5 & 8.2 & 17.59 & 0.20 \\
\hline
\end{tabular}
\end{center}
\end{table}

\begin{figure}[tbh]
\centering
  \includegraphics[width=0.9\hsize]{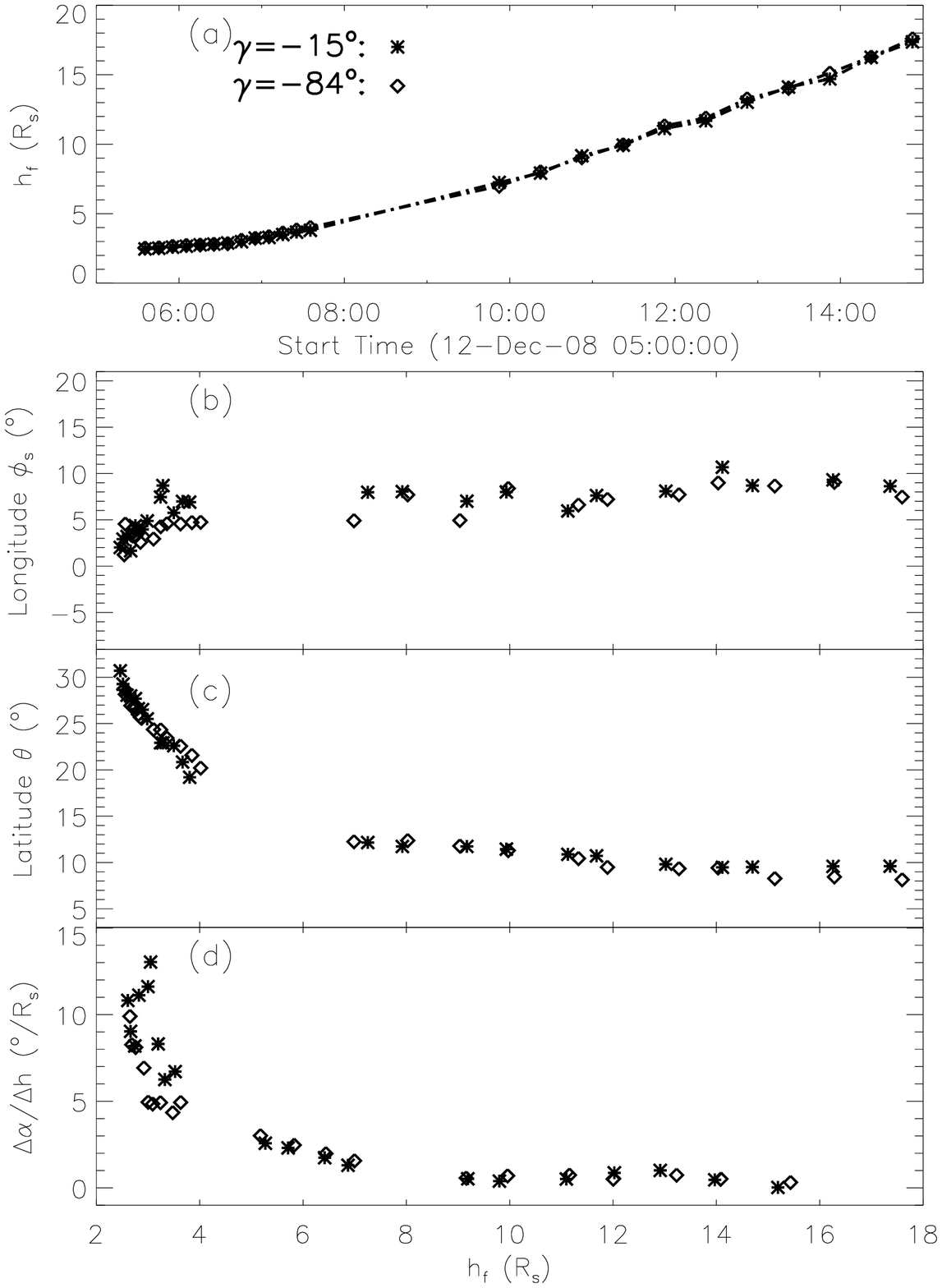}
  \caption{Comparison between the kinetic evolutions of the CME-1 event which were derived by the GCS model with two different tilt angles. The asterisks and the diamonds present the GCS model with the tilt angles of $-15^\circ$ and $-84^\circ$,
respectively. Panels (a) to (d) show the height-time, longitude-height, latitude-height, and deflection rate-height curves, respectively.}\label{fg_18}
\end{figure}

\begin{figure*}[tbh]
\centering
  \includegraphics[width=\hsize]{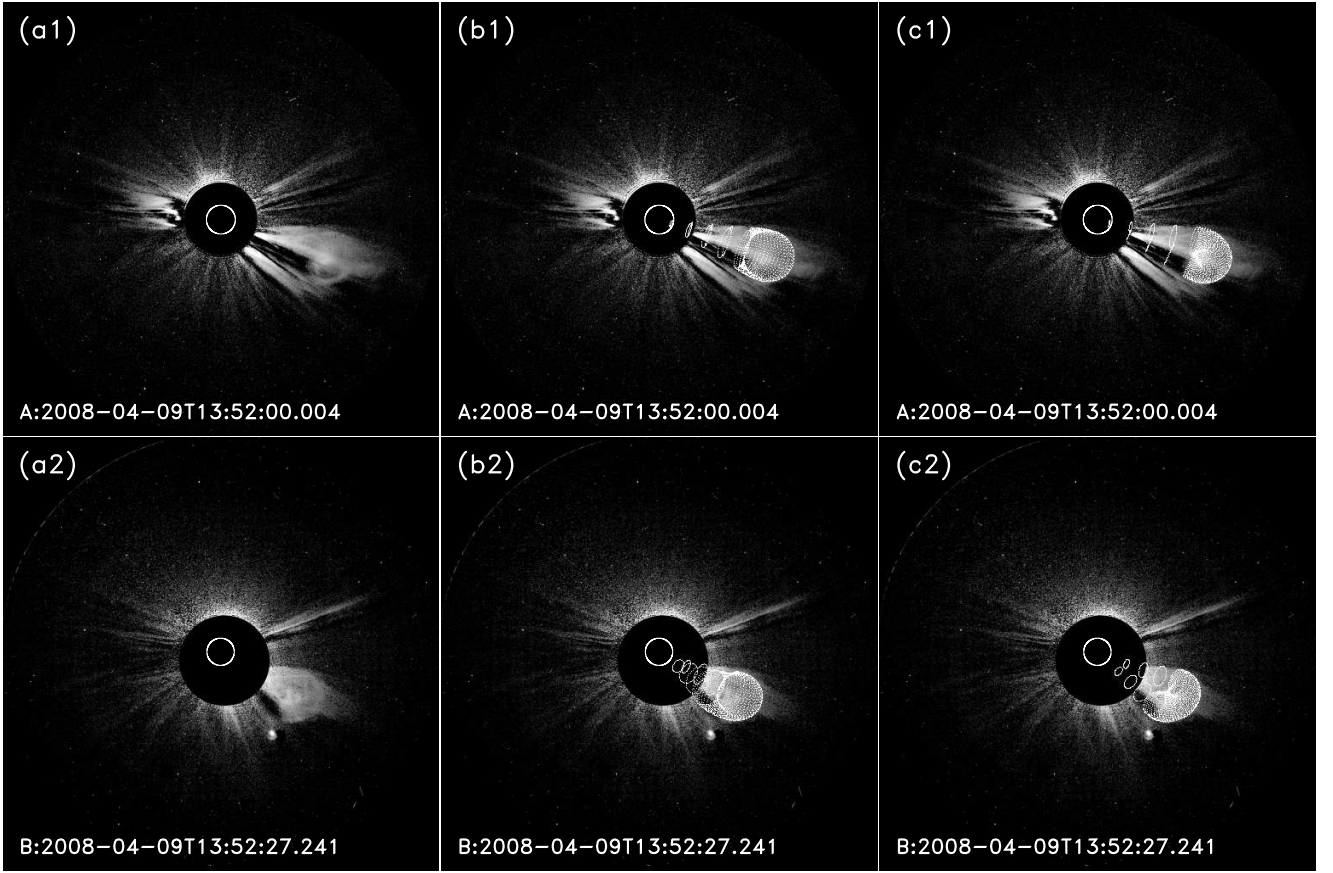}
  \caption{The fitting example of CME-2 with the GCS model of different tilt angles. From (a) to (c): The original CME images, the modeled wireframe images with the tilt angle of  $8^\circ$ and $-39^\circ$. The top and bottom panels present the results based on the STA and STB data, respectively.}\label{fg_19}
\end{figure*}

\begin{table}[tb]%\vskip -100pt
\begin{center}
%\tabcolsep 2pt
%\footnotesize %\centering
\caption{ The fitted free parameters of the 9 April 2008 CME derived by the GCS model with the tilt angle of $-39^\circ$ and
the half angle of $10^\circ$.}\label{tb_07}
\begin{tabular}{cccccc}
\hline
Time & $\phi_c$ &$\phi_s$ &$\theta$ \\
$[UT]$ & $[deg]$ & $[deg]$ & $[deg]$
& \raisebox{1.6ex}[0pt]{$h_f/R_s$} & \raisebox{1.6ex}[0pt]{$\kappa$} \\
\hline \multicolumn{6}{c}{COR1} \\
10:45 & 186.7 & 95.8 & -20.8 & 2.36 & 0.11 \\
10:55 & 190.7 & 99.8 & -20.6 & 2.56 & 0.11 \\
11:05 & 191.2 & 100.4 & -20.1 & 2.79 & 0.11 \\
11:15 & 191.7 & 101.1 & -20.2 & 3.04 & 0.11 \\
11:25 & 193.2 & 102.6 & -19.4 & 3.32 & 0.11 \\
\multicolumn{6}{c}{COR2} \\
13:22 & 197.5 & 108.0 & -19.7 & 8.57 & 0.14 \\
13:52 & 197.1 & 108.0 & -19.5 & 9.84 & 0.14 \\
14:22 & 199.8 & 111.8 & -20.0 & 11.27 & 0.14 \\
14:52 & 199.8 & 111.1 & -20.5 & 12.73 & 0.14 \\
\hline
\end{tabular}
\end{center}
\end{table}

\begin{figure}[tbh]
\centering
  \includegraphics[width=0.9\hsize]{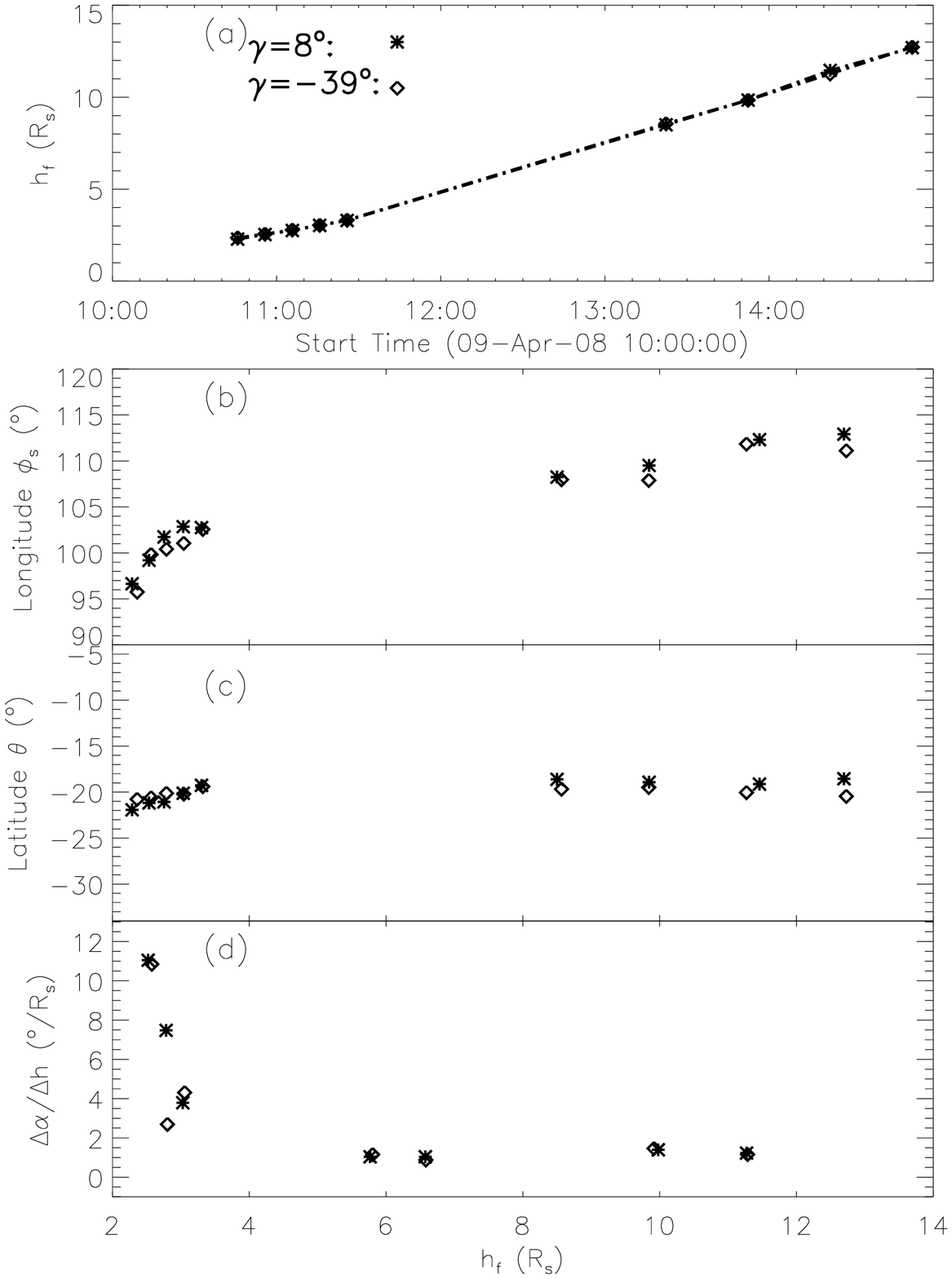}
  \caption{Comparison between the kinetic evolutions of the CME-2 event which were derived by the GCS model with two different tilt angles. The asterisks and diamonds present the GCS model with the tilt angles of $8^\circ$ and $-39^\circ$, respectively.
Panels (a) to (d) show the height-time, longitude-height, latitude-height, and deflection rate-height curves, respectively.}\label{fg_20}
\end{figure}

In this paper, the deflections of ten CMEs which occurred from November 2007 to the end of 2008 were studied. With the aid of the GCS model, eight of these CMEs are found to be deflected during their propagation in the corona. The distribution of the coronal magnetic field extrapolated from the SOHO/MDI magnetic synoptic charts suggests that the CMEs tend to deflect to the region with lower magnetic energy density. It confirms the result of \citet{Shen_etal_2011}.

The further quantitative analysis here reveals that the deflections and the magnetic energy density gradients have a better consistency in the lower corona or in the region with a stronger gradient of the magnetic energy density. The comparison of the deflection rate to the CME height and the speed suggests that CMEs have higher deflection rates in the inner corona, generally below 4 $R_s$. There is a positive correlation between the CME deflection rate and the strength of the magnetic energy density gradient. A stronger gradient may cause a larger deflection rate. Meanwhile, the CME speed has a negative effect on the deflection rate. A faster event tends to have a slower deflection. It is due to the gradient force of the magnetic energy density acting on the fast event lasted much shorter than that acting on the slow CME.

The fixed tilt angle implies the hypothesis that the CME did not rotate during the period of interest. We realized that this hypothesis may not be true. But it does not affect our result, because, even if we adjust the value of the tilt angle, the key parameters we have derived, including the longitude, latitude, and height, do not significantly change as long as the GCS flux rope can fit the observed CME well. Listed below are two examples illustrating this issue.

In our study, the tilt angle of the CME-1 was fitted to
$-15^\circ$. When we adjust the tilt angle, it is found
that the GCS flux rope with the tilt angle of $-84^\circ$
also fits the observations well. Figure \ref{fg_17} compares the GCS model
result between the two tilt angles, and Table \ref{fg_06} listed the
parameters for the comparison with Table \ref{tb_01}.

At the tilt angle of $-15^\circ$ the GCS flux rope is in
axial-view, while at the tilt angle of $-84^\circ$ the
GCS flux rope is in side-view.  Although the tilt angles are very different, the key parameters do not have significant differences. The maximal differences in the height, longitude, and latitude between the two cases are about 0.2 $R_s$, $4^\circ$ and
$2^\circ$, respectively. The variations in the three parameters are also shown by the asterisks and diamonds in Figure \ref{fg_18}. It can be seen that the two symbols are almost overlapped.

This event was studied by \citet{Liu_etal_2010b}, in which the tilt
angle was chosen as $-53^\circ$. By comparing our results
with the parameters which derived by the GCS model with the tilt
angle of $-53^\circ$ for the data point recorded at 12:52
UT given in the paper of \citet{Liu_etal_2010b}, it is found that the
differences in longitude and latitude are both $2^\circ$
only.

Similar to the CME-1 event, we find that the GCS flux rope with the
tilt angle of $-39^\circ$ also fits the observed shape of
CME-2. The fitting results of the GCS model with two different tilt angles are presented in Figure \ref{fg_19} and listed in Table \ref{tb_07}. Although the tilt angles are different, the key parameters do not have significant differences. The maximal differences in the height, longitude, and latitude between the two different tilt angles are about 0.1 $R_s$,
$2^\circ$ and $2^\circ$, respectively. Also the evolutions in the height, longitude, and latitude of the CME under the two different conditions are quite similar, as shown by the asterisks and diamonds in Figure \ref{fg_20}.

In addition, six of our ten events were also listed in Table \ref{tb_01} of the paper by \citet{Thernisien_etal_2009}. By comparing our Table \ref{tb_05} with their table, we find that the difference between the longitudes is mostly within $4^\circ$ and four of the six events are just $2^\circ$, and the difference between the latitudes is less than $1^\circ$ except for one event, which is about $3^\circ$. It is noticed that the longitudinal difference for the data point recorded at 17:52 UT of the 5 April 2008 event is about $14^\circ$. By fitting the CME images with the parameters given by \citet{Thernisien_etal_2009}, we find that the difference is caused by the different front edge selection. Even if we adopted the CME front edge selected by them, and performed the same analysis, it can be found that the CME would manifest the same deflection behavior.

In summary, the CME deflection is mainly controlled by the gradient of the coronal magnetic field based on our statistical study. The results confirm that the theoretical method proposed by the \citet{Shen_etal_2011} is able to quantitatively describe the CME deflections. Moreover, we believe that the method can be developed into a promising model, magnetic energy density gradient (MEDG) model, of predicting the CME deflection in the corona, though the basic idea of it is very simple. In this model, the gradient of the magnetic energy density is treated as a major cause of the CME deflection. Actually, the polarity of the background magnetic field may also have some effect on the deflections of CMEs \citep{Chane_etal_2005,Isenberg_Forbes_2007}. Besides, it should be noted that the gradient of the magnetic energy density decreases quickly with increasing height. When a CME propagates outward, the gradient of the background magnetic field may become weak rapidly. Such weak gradient would not be sufficient to make a CME deflected obviously, particularly during the propagation of a CME in the interplanetary space. This implies that there should be another mechanism to cause the CME deflection in the interplanetary space, which had been reported by \citet{Poomvises_etal_2010} and \citet{Lugaz_etal_2010}. A possible candidate mechanism is the CME's interaction with the background solar wind as proposed by \citet{Wang_etal_2004b,Wang_etal_2006c}.

\acknowledgments{We acknowledge the use of the data from STEREO/SECCHI and SOHO/MDI. We also acknowledge the use of the GCS model that developed by A. Thernisien. We are grateful to the anonymous referee for his/her kindly and constructive comments. This work is supported by grants from 973 key projects 2011CB811403, NSFC 40874075, 40904046, FANEDD 200530, CAS KZCX2-YW-QN511, 100-Talent program of CAS, and the fundamental research funds for the central universities. }

\bibliographystyle{agufull}
\bibliography{ahareference}
%\end{article}
\end{document}